

\magnification=\magstep1
\vsize=22.7 truecm
\hsize=16.5 truecm
\baselineskip=15pt
\tolerance=5000

\font\grand=cmbx10 at 14.4truept

\rightline{ UdeM-LPN-TH-92/103 \break}
\rightline{ CRM-1832 \break}
{}
\vskip 1truecm

\def\A{{\cal A}}
\def\g{{\cal G}}
\def\la{\langle}
\def\ra{\rangle}
\def\tr{{\rm tr\,}}
\def\La{\Lambda}
\def\ellg{\ell(gl_n)}
\def\Tr{{\rm Tr\,}}
\def\pa{\partial}
\def\th{\theta}
\def\res{{\rm res\,}}
\def\LPR{{\hat L}_1^{1/r}}
\def\dhu{{{\delta {\cal H}}\over {\delta u}}}
\def\dhw{{{\delta {\cal H}}\over {\delta w}}}
\def\bu{{\overline u}}
\def\bw{{\overline w}}
\def\sqr#1#2{{{\vbox{\hrule height.#2pt
      \hbox{\vrule width.#2pt height#1pt \kern#1pt
        \vrule width.#2pt}
      \hrule height.#2pt}}}}
\def\square{\mathchoice\sqr34\sqr34\sqr{2.1}3\sqr{1.5}3}
\centerline
{\grand Generalized Drinfeld-Sokolov Reductions and KdV Type Hierarchies}

\vskip 1.5truecm
\centerline{
L. Feh\'er${}^{1,3,}$\footnote*{
On leave from Bolyai Institute of Szeged University, H-6720 Szeged, Hungary.},
J. Harnad${}^{2,3}$
and I. Marshal${\rm l}^{2,3,}$\footnote{**}{
Present address:
Department of Mathematics, Leeds University, Leeds LS2 9JT, U.K.}
}
\bigskip
\centerline{\it ${}^{1}$ Laboratoire de physique nucl\'eaire}
\centerline{\it
Universit\'e  de Montr\'eal, C.~P. 6128, Montr\'eal, Canada H3C 3J7}

\medskip
\centerline
{\it ${}^{2}$ Department of Mathematics and Statistics}
\centerline{\it
Concordia University, 7141 Sherbrooke W., Montr\'eal, Canada H4B 1R6}

\medskip
\centerline{\it ${}^{3}$ Centre de recherches math\'ematiques}
\centerline{\it
Universit\'e de Montr\'eal, C.~P. 6128-A, Montr\'eal, Canada H3C 3J7}

\vskip 1.5truecm
\centerline{\bf Abstract}
\vskip 0.4truecm
Generalized Drinfeld-Sokolov (DS)  hierarchies are  constructed through
local reductions of Hamiltonian flows generated by monodromy invariants on
the dual of a loop algebra. Following earlier work of De Groot {\it et al},
reductions based upon graded regular elements of arbitrary Heisenberg
subalgebras are considered. We show that, in the case of the nontwisted
loop algebra $\ell(gl_n)$, graded regular elements exist only in those
Heisenberg subalgebras which correspond either to the partitions of $n$ into
the sum of equal numbers $n=pr$ or to equal numbers plus one $n=pr+1$.
We prove that the reduction belonging to the grade $1$ regular
elements in the case $n=pr$  yields the $p\times p$ matrix version of the
Gelfand-Dickey $r$-KdV hierarchy, generalizing the scalar case $p=1$
considered by  DS. The methods of DS are utilized throughout the analysis,
but formulating the reduction entirely within the Hamiltonian framework
provided by the classical r-matrix approach leads to some simplifications
even for $p=1$.

\vfill\eject

\bigskip
\centerline{\bf 0. Introduction }
\medskip

The generalized KdV type hierarchies of Drinfeld and Sokolov
(DS) are among the most important examples in the
field of integrable evolution equations [1].
They also play an important r\^ole in current studies of
two-dimensional gravity [2] and in conformal field theory [3].
The ``second Gelfand-Dickey'' Poisson bracket of these
bihamiltonian systems is
a reduction of the affine current algebra Lie-Poisson bracket,
and it gives an extension
of the Virasoro algebra by conformal tensors.
Such extended conformal algebras are called
${\cal W}$-algebras and have received a lot of
attention recently [4-6].

The motivation for the present work was to gain,
from a purely Hamiltonian viewpoint,
a better understanding of the reduction procedure used in [1]
and the generalizations proposed in a recent series of papers [7-9]
aimed at the construction of new integrable
hierarchies and ${\cal W}$-algebras.

In the DS construction [1]
one starts by considering a
first order matrix differential operator of the form
$$
{\cal L}=\partial +\mu\,,
\qquad\hbox{with}\qquad
\mu =q + \Lambda \ ,
\eqno(0.1)$$
where $\Lambda$ is a matrix representing a
grade $1$ regular element
of the principal Heisenberg subalgebra of a loop
algebra, and $q$ is a  smooth mapping from $S^1$ into an
appropriate subspace of the loop algebra
represented by ``lower triangular matrices''.
The crucial step of the construction is to transform
$\mu$ into the Heisenberg subalgebra by a conjugation of ${\cal L}$.
That this can be achieved by an algebraic, recursive procedure is due to
the fact that the grades in $q$ are lower than the grade of $\Lambda$
and $\Lambda$ is regular.
The fact that $\Lambda$ is regular also implies that
the stabilizer of the transformed operator is given
by the Heisenberg subalgebra.
The compatible zero curvature equations
are obtained
from the positively graded generators of this Heisenberg
subalgebra by transforming them into the
stabilizer of ${\cal L}$ and applying a splitting
procedure.
The system exhibits a gauge invariance
under a nilpotent, ``strictly lower triangular'',
gauge group.

The authors of [7] realized (see also [10], [11]) that the DS
construction can be applied in more general circumstances.
They proposed to derive new hierarchies
by replacing $\Lambda$ in the above by
{\it any} positive, graded regular element
of {\it any} graded Heisenberg subalgebra,
and correspondingly
modifying the DS  definition of the variable
$q$ and the gauge group.
However,
apart from some very simple examples,
they did not investigate which systems can be
obtained on the basis of this rather general proposal.
We shall see here that the number of new hierarchies arising
from this approach,
which in [7] were termed  type I hierarchies,
is in fact rather limited, since graded regular elements
do not exist in most
 Heisenberg subalgebras.
For simplicity, we shall investigate here the
case of the nontwisted loop algebra based on the
general linear Lie algebra $gl_n$, in which case the
inequivalent Heisenberg subalgebras are classified
by the partitions on $n$ [12].
By using
the explicit description of the inequivalent Heisenberg
subalgebras given in [13], we shall prove
that graded regular
elements exist only for the special partitions
$$
n=pr
\qquad {\rm and} \qquad n=pr+1 .
\eqno(0.2)$$

After explaining this observation,
we shall give a detailed analysis of the first
series of ``nice cases'' under (0.2).
The grade $1$ regular elements of the corresponding Heisenberg subalgebra
are the $n\times n$ matrices of the form
$\Lambda_{r,p}=\Lambda_r\otimes D$, where $D$ is a $p\times p$ diagonal
matrix such that $D^r$ has distinct, non-zero eigenvalues,
and $\Lambda_r$ is the usual $r\times r$
``Drinfeld-Sokolov matrix'' containing $1$'s above the diagonal and
the spectral parameter $\lambda$ in the lower-left corner.
The ``constrained manifold'' of the generalized
DS reduction will be taken to be the space of ${\cal L}$'s of the
form (0.1), where now $\Lambda=\Lambda_{r,p}$ and
$q$ is a mapping from $S^1$ into the block lower triangular
subalgebra of $gl_n$, with $p\times p$ blocks.
As in the $p=1$ case considered by DS,
the ``reduced space'' will be obtained by factorizing this
constrained manifold by the group of nonabelian gauge transformations,
$$
{\cal L}\rightarrow g{\cal L}g^{-1}\,,
\eqno(0.3)$$
where $g$ is now block lower triangular, having $p\times p$
unit matrices in the diagonal blocks.
We shall place the construction in a Hamiltonian setting
from the very beginning, from which it will be clear
that the reduced space is a bihamiltonian manifold that carries
the commuting hierarchy of Hamiltonians provided by the monodromy
invariants of ${\cal L}$.
It will also be clear that the locality of the
reduced system is guaranteed by construction.

In order to describe the reduced system
(i.e.,~the generalized KdV hierarchy) in terms of gauge invariant variables,
we shall prove the following facts, which
extend the $p=1$ results of [1].
First, the reduced space is the space
of ``matrix Lax operators'' of the form
$$
L=(-D)^{-r} \partial^r + u_1 \partial^{r-1}+\ldots
+u_{r-1}\pa +u_r\ ,
\eqno(0.4)$$
where the $u_i$ are now smooth mappings
from $S^1$ into the space of $p\times p$ matrices.
Second, the reduced bihamiltonian structure is that
given by the two compatible (matrix) Gelfand-Dickey Poisson brackets;
the second Poisson bracket algebra qualifies as a
classical ${\cal W}$-algebra.
Third, the hierarchy of commuting flows is
generated by the Hamiltonians
given by integrating the componentwise
residues of the fractional (including integral) powers of
the pseudo-differential operators obtained by
diagonalizing the matrix Lax operators.
In short,
the DS reduction extends in this case to yield the $p\times p$ matrix
version of the well-known (e.g.~[14]) Gelfand-Dickey $r$-KdV hierarchy.

The main additional step required in computing the Hamiltonians
in the $p\times p$ matrix case consists in the diagonalization
of $L$.
Those Hamiltonians which are obtained from
the {\it integral} powers of the diagonalized Lax operators
generate bihamiltonian ladders
whose first elements are the Hamiltonians
given by integrating the diagonal components of $u_1$,
which are Casimirs of the first Poisson structure.
The number of these bihamiltonian ladders
(which are missing in the scalar case)
is $p-1$ since the integral of $\tr(D^r u_1)$ is a Casimir
of both Poisson structures.
The other Hamiltonians can also be described
as integrals of trace-residues of independent
fractional powers of $L$, without diagonalization.

The KdV type hierarchies based on matrix Lax operators
of the type (0.4) have been investigated before in
refs.~[15-17] and more recently in [18].
In refs.~[15-17] the
matrix $(-D)^{-r}$ (i.e.,~the coefficient of the leading
term of $L$)
was required to be regular because
this implies
the existence of the maximal number of independent
$r$th roots of $L$
and corresponding commuting flows.
It is interesting to see this condition re-emerge here
from requiring that the matrix $\Lambda$ used in the
reduction procedure be a regular element of the Heisenberg subalgebra.
In these papers, the additional assumption
was made that the diagonal part of $u_1$ vanishes.
Setting $[u_1]_{\rm diag}=0$
is consistent with the equations of the
hierarchy resulting from the DS reduction
and in fact corresponds to an additional Hamiltonian
symmetry reduction.
(See Remarks 2.5 - 2.7.)

The recent preprint [18] deals with
the hierarchy defined by the
fractional powers of ``covariant Lax operators'',
which are equivalent to operators of the form (0.4).
More precisely,  the case of the Lie algebra $sl_n$
was considered,
which in our approach corresponds to imposing the constraints
${\rm tr} (D^r u_1)=0$.
However, instead of taking a regular
 matrix for $(-D)^{-r}$, the unit matrix was used.
 As a result, the hierarchy obtained
 is much smaller than the one following from  the
DS reduction
 using a regular element.

To emphasize the
link between the approach used in [1], [7] and
that based upon the Adler-Kostant-Symes (AKS)
construction as presented in the present work,
we give the following elementary lemma.

\smallskip
\noindent
{\bf Lemma 0.1.} {\it
Let $\A$ be a Lie algebra.
Let $\mu_0\in \A^*$ be given and
$X\in {\rm cent}({\rm stab}(\mu_0)) \subset \A$ be an
element in the center of its stabilizer.
Then there exists an element
$\varphi\in {\cal I}(\A^*)$ of the ring of
${\rm ad}^*$-invariant functions on
$\A^*$ such that $\nabla \varphi\vert_{\mu_0} = X$, where, as usual, an
identification has been made between the cotangent space to $\A^*$ at
$\mu_0$ and $(\A^*)^* \sim \A$.
Conversely, if $\varphi\in
{\cal I}(\A^*)$, then $\nabla\varphi\vert_{\mu_0} := X \in
{\rm cent}({\rm stab}(\mu_0))$.
}
\smallskip

The point of this lemma is that, when applied to the loop algebra
$\A=\ell({\cal G})$ of a Lie algebra ${\cal G}$, with the
splitting $\A = \A_+ + \A_-$ into the sum of subalgebras consisting of
positive and negative powers in the loop parameter $\lambda$,
it implies that,  starting from any $X\in {\rm cent}({\rm stab}(\mu_0))$,
with $\mu_0 \in ({\cal A}_-)^*\sim ({\cal A}^*)_+$,
the flow induced in $({\cal A}_-)^*$ by exponentiation
and factorization in the group is given by integration of a Lax type equation
$$
{d\mu \over dt} =\pm ({\rm ad\,}^* {({\nabla \varphi}\vert_\mu)}_\pm) (\mu)
\eqno(0.5)$$
with $\varphi\in {\cal I}(\A^*)$.
Taking the Lie algebra ${\cal G}$ itself as a centrally extended
loop algebra in the space variable $x$, eq.~(0.5) becomes a
zero-curvature (Zakharov-Shabat) equation.
The commutativity of such flows is part of the
AKS theorem.

The above lemma
 underlies the equivalence
between the DS approach [1], [7], which is
based essentially upon
${\rm cent}({\rm stab}(\mu))$,
and the AKS approach, based on ${\cal I}(\A^*)$.
This equivalence is certainly known to specialists,
but in this paper it will be taken as the starting point and
{\it all} results will be derived from the AKS Hamiltonian
point of view.

This paper is organized as follows.
In Sec.~1 we collect results that are relevant
for understanding the DS type construction of compatible
zero curvature equations in a Hamiltonian setting.
In particular,  in Sec.~1.1 we recall
the relevant aspects of the AKS
construction of commuting flows.
In Sec.~2.2 we discuss a sufficient
condition that can be used to
obtain local ${\rm ad}^*$-invariant
Hamiltonians from the asymptotic expansion of the
monodromy matrix of an appropriate first order
matrix differential operator.
These two sections naturally lead us to look for
the graded regular elements of the Heisenberg
subalgebras of a loop algebra as the starting point
for obtaining generalized DS hierarchies.
The solution to this problem  is given
in the case of the nontwisted loop algebra $\ell(gl_n)$
by Theorem 1.7 in  Sec.~1.3.
Sec.~2 is devoted to the generalized DS reduction yielding
the matrix Gelfand-Dickey hierarchy.
The description of the reduced space
is established in Sec.~2.1,
the Hamiltonian structures are described in Sec.~2.2
and the Hamiltonians themselves are given in Sec.~2.3.
The Poisson brackets and the first few Hamiltonians
are computed explicitly for an example in Sec.~2.4.
The main results of Sec.~2 are Theorem 2.4
which identifies the reduced Poisson structures as the first
and second Gelfand-Dickey Poisson structures,
and Corollary 2.11 of Theorem 2.10
which gives the generating set of commuting Hamiltonians.
The paper concludes with remarks relating the
matrix Gelfand-Dickey hierarchies to nonabelian
``conformal'' and ``affine'' Toda systems,
comments on ${\cal W}$-algebras and on the case $n=pr+1$,
and further remarks concerning the literature
and some open problems.

\vfill\eject

\centerline{\bf 1. The AKS construction and local reductions}
\medskip

In the first two sections we review some well-known results
about the AKS approach in loop algebras.
We shall naturally be led to considering the problem of finding
all the graded regular elements of the Heisenberg subalgebras
of $gl_n\otimes C[\lambda,\lambda^{-1}]$, which are given in
Section 2.3.

\bigskip

\noindent
{\bf 1.1. The AKS construction}
\medskip

Here we summarize those points of the AKS
(or r-matrix) approach which we shall need.
Readers unfamiliar with the construction could consult,
for example, refs.~[19-21] for further details.

Let $\A$ be a Lie algebra with Lie bracket $[\ ,\ ]$.
Suppose that $R$ is a
classical r-matrix; that is, $R\in {\rm End\,}\A$ and the bracket
$[X,Y]_R: \A \times \A \to \A\ $ given by
$$
[X,Y]_R ={1\over 2}[RX,Y]+{1\over 2}[X,RY]
\eqno(1.1)
$$
is also a Lie bracket.
For a function $\varphi$ in $C^\infty(\A^*)$ we define
its differential
$\nabla_\alpha \varphi \in {\cal A}$
at a point  $\alpha \in {\cal A}^*$
by
$$
{d\over dt} \varphi (\alpha +t \beta ){{\vert}_{t=0}}=
\la \beta , \nabla_\alpha \varphi \ra
\qquad \forall \beta \in {\cal A}^*\,,
\eqno(1.2)
$$
where $\la\ ,\ \ra$ is the dual pairing.
Let
$$
I(\A^*) =
\{\,\varphi\in C^\infty(\A^*)\, \vert\,
0 =
\la ({\rm ad}^* X)(\alpha ), \nabla_\alpha \varphi \ra
=\la \alpha , [\nabla_\alpha \varphi , X]\ra
\quad \forall X  \in \A\ \},
\eqno(1.3)
$$
be the set of ad$^*$-invariant functions on $\A^*$.
Note that in many cases we can think in terms of the ${\rm Ad}^*$-action on
$\A^*$
of a Lie group $G$ corresponding to $\A$, in which case
$$
I(\A^*)=\{\,\varphi\in C^\infty(\A^*) \,\vert\,
\varphi ({\rm Ad}^*_g \alpha )=\varphi (\alpha )\quad \forall g\in G\ \}.
\eqno(1.4)
$$
Here ad$^*$ means the action dual to the
original Lie bracket $[\ ,\ ]$
on $\A$;  to refer to the action of $\A$ on $\A^*$ dual to
that given by the bracket $[\ ,\ ]_R$, we write ad$_R^*$.

A surprisingly large number of integrable systems arise
as consequences of the following result [20],
which is the r-matrix version of the AKS theorem, and
whose proof is a direct application of (1.1) and (1.3).

\smallskip
\noindent
{\bf Proposition 1.1.} {\it The  elements of
$I(\A^*)$ are an involutive family in $C^\infty (\A^*)$ with respect
to the $R$ Lie-Poisson
bracket on $\A^*$. That is, if $\varphi, \psi \in I(\A^*)$ then
$$
\{\varphi, \psi\}_R(\alpha) \equiv \la \alpha ,
[\nabla_\alpha \varphi ,\nabla_\alpha \psi ]_R \ra =0\,.
\eqno(1.5)
$$
The dynamical equation generated by the Hamiltonian
$\varphi\in I(\A^*)$ through the
R Lie-Poisson bracket has the generalized Lax form}
$$
\dot \alpha=({\rm ad}^*{1\over 2}R \nabla_\alpha \varphi )(\alpha)\,.
\eqno(1.6)$$

Consider now the special case that will be of interest
in what follows (see [21]).
Let $\g$ be a Lie algebra and set
$\A=\ell(\g)=\g\otimes C[\lambda,\lambda^{-1}] =
\{ \sum_{i=r}^s X_i\lambda^i \vert\, X_i \in \g\}$.
Let $\A_+=\g\otimes C[\lambda]$, $\A_-=\g\otimes \lambda^{-1}C[\lambda^{-1}]$,
so that $\A =\A_++\A_-$. We let $P_\pm$ be the projection operators defined by
this splitting and set $R=P_+-P_-$. For  any $\eta\in C[\lambda,\lambda^{-1}]$
we define $\hat \eta : \A \to \A$ by $(\hat \eta X)(\lambda)=\eta(\lambda)
X(\lambda)$.
Then
$$
R_\eta := R\circ \hat \eta
\eqno(1.7)$$
defines a classical r-matrix for any $\eta$, and
the corresponding Lie-Poisson brackets for different $\eta$ are compatible.

We can identify
$\ell(\g)^*$ with $\ell(\g^*)=\g^*\otimes C[\lambda,\lambda^{-1}]$
by means of the pairing $\la\ ,\ \ra$ given by
$$
\la\alpha,X\ra :=(\alpha (X)){{\vert}_{-1}}\ ,
\quad\hbox{i.e.}\quad
\la \sum_i\alpha_i\lambda^i\,,\,\sum_j
X_j\lambda^j\ra:=\sum_{i+j=-1} \alpha_i(X_j)\,.
\eqno(1.8)$$

\medskip
\noindent
{\it Remark 1.1.}
An equivalent procedure would be to keep the r-matrix fixed
and define a dual pairing for each $\eta\in C[\lambda, \lambda^{-1}]$
by inserting a factor $\eta$ into the definition (1.8).
In view of this, we see that in this case there is no essential
difference between the r-matrix formulation and the original
AKS splitting procedure.
In what follows, the collection of commuting Hamiltonian systems
determined by the elements of $I(\ell({\cal G}^*))$,
together with the compatible Poisson brackets $\{\ ,\ \}_{R_\eta}$
will be referred to as the  {\it AKS system}.

\medskip

The ${\rm ad}^*$-action of $\ell(\g)$ on
$\ell(\g^*)$ is given
by pointwise evaluation in $\lambda$.
This implies that one can in general obtain
elements of $I(\ell(\g^*))$ in the following way.
Let $\varphi \in I(\g^*)$ be an invariant polynomial;
choose any element $\varrho \in C[\lambda,\lambda^{-1}]$.
Then $\varphi_\varrho \in C^{\infty}(\ell(\g^*))$, given by
$$
\varphi_\varrho (\alpha) :=
\left(\varrho(\lambda)\varphi(\alpha(\lambda))\right){{\vert}_{-1}}\,,
\qquad \alpha=\sum_i \alpha_i \lambda^i\in \ell(\g^*),
\eqno(1.9)
$$
is in $I(\ell(\g^*))$.
If $I(\g^*)$ is generated by polynomials $\varphi$ then
$I(\ell(\g^*))$ is generated by the
corresponding functions $\varphi_\varrho$.
The Hamiltonian vector fields determined by these
invariant functions on $\ell(\g^*)$ obey the following relations,
$$
\{ f\,,\,\varphi_{\varrho \chi}\}_{R_\eta}=\{ f\,,\,
\varphi_\varrho\}_{R_{\eta\chi }}\,,
\qquad
\forall\, f\in C^\infty(\ell(\g^*))\,,
\quad \forall\, \eta,\varrho,\chi\in C[\lambda,\lambda^{-1}].
\eqno(1.10)$$

Notice [21] that the space
$$
{\widehat {\cal M}}_{-m,n}:=\{\,\sum_{i=-m}^n u_i\lambda^i\,\vert
\,u_i\in g^*\ \}\subset \ell(\g^*)
\qquad {\rm where}\quad m \ge 0, \, n\ge -1,
\eqno(1.11)$$
is a Poisson subspace for any of the r-matrices $R_k:=R\circ \hat \eta_k$,
where $\eta_k(\lambda):=\lambda^k$, if $-m\leq k \leq n+1$.
Furthermore, if $k\leq n$ then $u_n$ is
a Casimir element; i.e.,~constant under any
Hamiltonian flow of the $R_k$ Lie-Poisson bracket.
Hence the affine subspace
${\overline {\cal M}}_{-m,n}\subset {\widehat {\cal M}}_{-m,n}$
having $u_n$ fixed has $n+m+1$ compatible Poisson brackets,
and the restriction of the elements of $I(l(\g^*))$
provides a commuting family of Hamiltonians
with respect to any of them.
Note that the same computation which shows
${\overline {\cal M}}_{-m,n}$ is a Poisson
subspace of $\ell(\g^*)$ justifies restricting ourselves
to $\ell(\g^*)$, which is strictly speaking only a subspace of
$\ell(\g)^*$. In both cases
one simply has to check that an arbitrary Hamiltonian flow
determined by
$$
\dot\alpha = ({\rm ad}^*_{{R_\eta}}\nabla_\alpha H)( \alpha )
\eqno (1.12)
$$
does not leave the space.

\smallskip

 From now on we take  $\g$ to be ${\widetilde {gl_n}}^\wedge$,
the central extension of
the algebra of smooth loops in $gl_n$, i.e.,
$\g=\{\,(X,a)\,\vert\,X: S^1\to gl_n\,,\, a\in {\bf C}\,\}$
with\footnote*{The periodic space variable parametrizing $S^1$
is usually denoted by $x\in [0,2\pi]$, and tilde
signifies here ``loops in $x$''.
}
$$
[(X,a),(Y,b)]=\left( XY-YX\,,\,\int_0^{2\pi} dx\,{\rm tr}X'(x)Y(x) \right)\ .
\eqno(1.13)
$$
As usual, we represent (a dense subspace of)
the dual space $\g^*$ as the space of first
order matrix differential operators
$$
{\cal L}=(e\partial + \mu)\longleftrightarrow (\mu ,e)\in \g^* \,,
\eqno(1.14)
$$
with $\mu\in \widetilde{gl_n}\sim {\widetilde{gl_n}}^*$, $e\in {\bf C}$
and dual pairing
$$
\la (\mu,e)\,,\,(X,a)\ra = ea+\int_0^{2\pi} dx\,
{\rm tr\,}\mu (x) X(x) \,.
\eqno(1.15)
$$
Upon introducing the spectral parameter $\lambda$,
the elements of $\ell(\g)$ and $\ell(\g^*)$ are given by pairs
$(X,a)$ and $(\mu,e)$, where
$X=\sum X_j \lambda^j$, $\mu=\sum \mu_i \lambda^i$
are now mappings from $S^1$ into $\ell(gl_n)$,
and $a=\sum a_j\lambda^j$, $e=\sum e_i\lambda^i$
are elements of $C[\lambda, \lambda^{-1}]$.
The ${\rm ad}^*$-action on $\ell(\g^*)$ is given by
$$
{\rm ad}^*(X,a) :
(\mu , e)\mapsto (X\mu -\mu X - eX', 0 )\ .
\eqno(1.16)
$$
The ${\rm Ad}^*$-action is given by  ``nonabelian gauge transformations'';
that is, by
$$
{\cal L}=e\partial +\mu
\mapsto g{\cal L} g^{-1}
=e\partial + (g \mu g^{-1} - e g' g^{-1})\,,
\eqno(1.17)
$$
where $g$ is an element of the loop group
$\ell({\widetilde{Gl}_n})$ associated to $\ell({\widetilde {{gl}_n}})$.

The ring of ${\rm ad}^*$-invariant functions
$I(\ell(\g^*))$ can be specified as follows.
Consider the linear problem
$$
\left(e\partial_x +\mu(x,\lambda)\right)\Phi(x,\lambda)=0\,,
\eqno(1.18)
$$
where $\Phi(x,\lambda)\in Gl_n$.
Then the eigenvalues of the monodromy matrix
$$
T(\lambda) = \Phi(2\pi,\lambda)(\Phi(0,\lambda))^{-1}\,,
\eqno(1.19)
$$
viewed as functions on $\ell(\g^*)$,
generate the ring $I(\ell(\g^*))$.

The number of invariants is infinite
due to the parametric dependence of $T$ on $\lambda$.
In general these are not local functionals of $\mu$;
i.e.,~they cannot be expressed
as functions of a finite number of
integrals of
local densities in the components
$\mu_i(x)$ of $\mu(x,\lambda) =\sum_i \mu_i(x) \lambda^i$
and their derivatives.
However, under certain conditions (see Sec.~1.2), it is possible,
using the asymptotic expansion of $T(\lambda)$, to determine
an infinite set of local, commuting Hamiltonians on appropriate
Poisson subspaces of $\ell({\cal G}^*)$,
or reductions thereof.

Following the above,
let us take $\eta$ in (1.7) to be of the form
$$
\eta(\lambda) = \eta_0 + \eta_1\lambda ,
\eqno(1.20)
$$
and consider the subspace
${\cal M}_{0,1}\subset{\overline {\cal M}_{0,1}}$
consisting of elements of the form
$$
(\mu,e)(\lambda) = (J + C_1\lambda,\, e_0 + e_1\lambda)\,,
\eqno (1.21)
$$
where $C_1\in gl_n$, $e_0, e_1\in {\bf C}$ are constants
and $J: S^1 \to gl_n$ is arbitrary.
Since this is a Poisson subspace of $\ell(\g^*)$
with respect to the $R_\eta$ Lie-Poisson bracket,
the Poisson bracket can
be restricted to functions that depend only on $J$, giving
$$ \eqalign{
\{\varphi, \psi\}_{R_\eta}(J) =&
-\eta_0\left(\int_{S^1}\tr C_1[
{{\delta \varphi}\over {\delta J}}, {{\delta \psi}\over {\delta J}} ] +
e_1\int_{S^1}\tr
\left({{\delta \varphi}\over {\delta J}}\right)'
{{\delta \psi}\over {\delta J}} \right) \cr
&+  \eta_1\left(\int_{S^1}\tr J
[{{\delta \varphi}\over {\delta J}} , {{\delta \psi}\over {\delta J}} ] +
e_0\int_{S^1}\tr
\left({{\delta \varphi}\over {\delta J}}\right)'
{{\delta \psi}\over {\delta J}}
\right)\ \, .\cr}
\eqno (1.22)
$$
The functional differential
${{\delta \varphi}\over {\delta J}} : S^1 \to gl_n$ is
defined here by the usual formula,
$$
{d\over dt}\varphi (J + t K ){{\vert}_{t=0}} =
\int_{S^1}  \tr K {{\delta \varphi}\over {\delta J}}
\qquad  \forall\,K : S^1\to gl_n\ .
\eqno(1.23)$$
 From now on we choose the
values of the Casimirs $e_0$ and $e_1$ to be $e_0 =1$ and $e_1 = 0$.
The space of Poisson brackets given by $(1.22)$ has a basis given
by the choices
$\eta(\lambda) = 1$ and by $\eta(\lambda) = \lambda$.
A common way of describing the above is to say
that the space of functions of $J$ has two compatible
Poisson brackets:
$$
\{\varphi, \psi\}_1 (J) =
-\int_{S^1}\tr C_1[{{\delta \varphi}\over {\delta J}} ,
{{\delta \psi}\over {\delta J}} ]
\eqno (1.24{\rm a})
$$
and
$$
\{\varphi, \psi\}_2 (J) =
\int_{S^1}\tr \left(J [{{\delta \varphi}\over {\delta J}} ,
{{\delta \psi}\over {\delta J}}  ] +
\left({{\delta \varphi}\over {\delta J}}\right)^\prime
{{\delta \psi}\over {\delta J}}\right)\,.
\eqno (1.24{\rm b})
$$
We have denoted
the Poisson bracket belonging to $\eta=\lambda^i$ as
$\{\ ,\ \}_{i+1}$ in order to
be consistent with the traditional terminology of KdV
systems later.
Note that the first Poisson bracket is
just the Lie derivative of the second Poisson bracket
with respect to the vector field
that generates translations of $J$ in the direction $-C_1$.

In general, ``interesting'' examples result from
appropriate further reductions of a space ${\cal M}_{0,1}$.
A key observation in this respect is that the group
consisting of those $\lambda$-independent nonabelian gauge transformations
$g:S^1\to Gl_n$ for which
$$
g C_1 g^{-1}=C_1
\eqno(1.25)
$$
is a {\it symmetry group} of the AKS system restricted to
${\cal M}_{0,1}$.
Indeed the transformations of
${\cal M}_{0,1}$ obtained from (1.17) by using
this group preserve both the compatible Poisson structures
and the monodromy invariants.
Thus one can use this group or any of its subgroups when
searching for ``nice symmetry reductions''
of the AKS system.

\medskip
\noindent
{\it Remark 1.2.}
The apparent generalization obtained
by considering ${\overline {\cal M}}_{-m,n}$ with
$n+m>1$ instead of ${\overline {\cal M}}_{0,1}$
does not add any new interesting structure,
since the reduction only affects
$u_{n-1}$ and $u_n$; the other terms remain generic.

\bigskip
\noindent
{\bf 1.2. How to obtain local invariants}
\medskip

Let us consider the operator ${\cal L}=(\partial+\mu)$ where
$\mu: S^1\to \ell(gl_n)$.
There is a fairly well-known sufficient
condition  on the form of the function $\mu$
that one can impose in order
to guarantee the locality of the monodromy invariants
of ${\cal L}$.
This condition involves the
{\it graded regular} elements of the affine algebra
$\ell(gl_n)$, and  we shall explain a variant of it below.

Consider  a fixed element $\La \in \ell(gl_n)$ and
denote its kernel and image in the adjoint representation by
$$
{\cal K}={\rm Ker\,}({\rm ad\,}\Lambda)
\qquad {\rm and}\qquad {\cal I}={\rm Im\,}({\rm ad\,}\Lambda)\ .
\eqno(1.26)$$
Of course, ${\cal K}$ is a subalgebra of $\ellg$.
For $\La$ a {\it regular} element one has
$$
\ellg = {\cal K}+{\cal I}\,,\qquad
{\cal K}\cap {\cal I}=\{ 0\}\,,\qquad
{\cal K}:\ {\rm abelian}\ .
\eqno(1.27)$$
We are interested in regular elements that are {\it homogeneous
with positive degree} with respect to some integral grading.
An integral grading of $\ellg$ can be
defined by the eigenspaces of a
linear operator
$d_{N,H}:\ellg\rightarrow \ellg$
of the form
$$
d_{N,H}=N\lambda {d\over d\lambda} + {\rm ad\,} H\, ,
\eqno(1.28)$$
where $N$ is a non-zero integer and $H$
is a diagonalizable element of $gl_n$ with
integral spectrum in the adjoint representation.
This formula in fact defines a
derivation on $\ellg$ with integral eigenvalues
and finite dimensional eigenspaces.

The following proposition, which generalizes the corresponding
well-known result for the homogeneous grading case  ([1], [19], [22]),
 states the existence
of a solution of the linear problem
given by a series that can be computed
by an algorithm involving
only linear algebraic operations and integrations.

\medskip
\noindent
{\bf Proposition 1.2.}
{\it Let a grading (1.28) of $\ellg$  be given.
Let $\La$ be
a regular homogeneous element
of grade $l>0$.  Consider a
function $\mu : S^1 \to \ellg$ of the form
$$
\mu(x) = (q(x) +\Lambda)\,,
\eqno(1.29{\rm a})$$
where
$$
q(x)=\sum_{k<l} q^k(x) \qquad{\it with}\qquad
d_{N,H}(q^k)=k q^k\,.
\eqno(1.29{\rm b})$$
Then the linear problem
$$
(\partial_x + q(x) + \Lambda )\Phi(x)=0
\eqno(1.30)$$
has a unique solution of the form
$$
\Phi(x) = (I+W(x))e^{F(x)}(I+W(0))^{-1}\Phi(0)\,,
\eqno(1.31)$$
where
$$
F(x)\in {\cal K}\,,\quad W(x)\in {\cal I}\,,
\quad {\it and}\quad W(x)=\sum_{k<0}W^k(x)
\quad {\it with}\quad d_{N,H}(W^k)=kW^k\,.
\eqno(1.32)$$
Here the $W^k$'s are uniquely determined
differential polynomials in the components
of $q$ and $F$  is given by
$$
F(x)= -\int_0^x dy\, [q_{\cal K}(y) + (q(y)W(y))_{\cal K}+\Lambda]\ ,
\eqno(1.33)$$
where the subscript ${\cal K}$ refers to the
${\cal K}$ component in the decomposition}
$\ellg={\cal K}+{\cal I}$.

\medskip
\noindent
{\bf Corollary 1.3.} {\it The  monodromy matrix
of ${\cal L}=(\partial + q+\Lambda)$ is conjugate to
$$
\exp\bigl(F(2\pi)\bigr)
=\exp \bigl(-\int_0^{2\pi}dx
[q_{\cal K}(x)+(q(x)W(x))_{\cal K}+\Lambda]\bigl)\ ,
\eqno(1.34)$$
and thus its invariants are functions of
integrals of local densities in $q(x)$.}

\medskip
\noindent
{\it Proof.}
The procedure is essentially the same as given
in a special case in [22].
By substituting the ansatz
$$
\Phi = (I+W)e^F \Psi\,,
\eqno(1.35)$$
where $I$ is the unit matrix and $\Psi$ is a constant,
into (1.30) we obtain
$$
W^\prime +(I+W)F^\prime+(q+\Lambda)(I+W)=0\,.
\eqno(1.36)$$
If we decompose this equation according to
$\ellg={\cal I}+{\cal K}$ by using (1.32) together with
${\cal I}{\cal K}\subset {\cal I}$ and
${\cal K}{\cal K}\subset {\cal K}$, then the
${\cal K}$-component gives
$$
F^\prime +[q(I+W)]_{\cal K}+\Lambda =0\,,
\eqno(1.37)$$
which (up to a constant) can be integrated to give (1.33),
since ${\cal K}$ is abelian.
By substituting  (1.37) into (1.36), the ${\cal I}$-component
gives
$$
[\Lambda\,,\,W]+W^\prime -W(q+qW)_{\cal K}+(q+qW)_{\cal I}=0\,.
\eqno(1.38)$$
One can solve this equation recursively for the $W^k$'s
by using the grading assumptions of (1.29) and (1.32)
together with the fact that
${\rm ad\,}{\Lambda}$ maps ${\cal I}$ to ${\cal I}$
in a one-to-one manner since $\Lambda$ is regular.
This procedure obviously yields the $W^k$'s as differential polynomials
in $q$.
Finally, the integration constant $\Psi$ in (1.35) is fixed
by the initial condition, giving (1.31). \quad $\square$

Note that the recursive procedure appearing in the proof,
combined with the diagonalization of the generators of ${\cal K}$,
is also useful for computing the monodromy invariants in practice.
It should also be noted that in the above we have not considered
the convergence of the  series solution at all.
It is well known that such series do not converge
in general, and are to be considered as
asymptotic expansions in $\lambda$
(or alternatively as formal series).


\bigskip
\noindent
{\bf 1.3. The list of graded regular elements}
\medskip

We have seen that the graded regular elements of
$\ellg$ can be used to impose constraints
on the form of ${\cal L}=(\pa+\mu)$ leading to
local monodromy invariants.
The suggestion of De Groot {\it et al}
[7] was to use the graded regular elements of the
inequivalent graded Heisenberg subalgebras of the loop algebras [12]
to construct generalizations of the Drinfeld-Sokolov hierarchies.
The graded Heisenberg subalgebras of the loop algebra $\ellg$
(maximal abelian subalgebras that acquire a central
extension in $\ellg^\wedge$) have been given an
explicit description recently in [13],
where the authors were interested in the related vertex
operator constructions.
By using this description, we shall
show that graded regular elements
exist only in some exceptional Heisenberg subalgebras.
The complete list is given by Theorem 1.7 at the end of this section.

The graded Heisenberg subalgebras of $\ellg$ can be associated
to the {\it partitions} of $n$ in the following way [13].
First, for $m$ any natural number, we define the following
$m\times m$ matrices:
$$\eqalignno{
\La_m&=\lambda e_{m,1}+\sum_{k=1}^{m-1}e_{k,k+1}\ ,\cr
H_m&={\rm diag\,}[j,(j-1),\ldots,-(j-1),-j] \ ,
\qquad j={{m-1}\over 2}\ ,&(1.39)\cr
\noalign{\hbox{and}}
\sigma_m&=\exp [2\pi i H_m/m]\ ,}
$$
where $e_{i,j}$ is the standard elementary matrix with entry
$1$ in the $ij$th place and $0$ elsewhere.
Let a partition of $n$ be given by
$$
n=n_1+n_2+\cdots +n_k\ ,
\qquad \hbox{where}\qquad
n_1\geq n_2\geq \cdots \geq n_k\geq 1\ .
\eqno(1.40)$$
We associate to this partition the $n\times n$ matrix
$$
\sigma = {\rm diag\,}[\sigma_{n_1},\sigma_{n_2},\ldots,\sigma_{n_k}]\ ,
\eqno(1.41{\rm a})$$
and denote by $N$ the order of the inner automorphism of $gl_n$ acting
through conjugation by $\sigma$.
If we let
$$
N^\prime = {\rm lcm}(n_1,n_2,\ldots,n_k)\ ,
\eqno(1.41{\rm b})$$
we have
$$
N=\cases{N^\prime, &if $N^\prime ({1\over n_i}+{1\over n_j})$
          is even for all $i$,$j$;\cr
         2N^\prime,&if $N^\prime ({1\over n_i}+{1\over n_j})$
           is odd for some $i$,$j$.}
\eqno(1.41{\rm c})$$
We then introduce the $n\times n$ diagonal matrix $H$ via the
equation
$$
\sigma = \exp[2\pi i H/N]\ ,
\eqno(1.41{\rm d})$$
and consider the grading of $\ellg$ given by the eigenvalues of
$$
d= N\lambda {d\over d\lambda }+{\rm ad\,} H\ .
\eqno(1.42)$$
The Heisenberg subalgebra corresponding to the partition (1.40)
is spanned by the $n\times n$ ``block-diagonal''
matrices $\La$ of the following form:
$$
\La=\left[\matrix{ y_1 \La_{n_1}^{l_1}&{}&{}&{}\cr
              {}&y_2\La_{n_2}^{l_2}&{}&{}\cr
              {}&{}&\ddots&{}\cr
             {}&{}&{}&y_k\Lambda^{l_k}_{n_k}\cr}\right]\ ,
\eqno(1.43)$$
where the $l_i$ ($i=1,2,\dots,k$) are arbitrary integers and the $y_i$
are arbitrary  numbers.
This maximal abelian subalgebra of $\ellg$ is invariant under the
grading operator (1.42).
An  element $\Lambda$ of the form given by (1.43) is {\it regular} if
${\cal K}={\rm Ker}({\rm ad\,} \Lambda)$
is exactly the Heisenberg subalgebra (and not a larger space).
We next investigate the existence of the graded regular elements for some
simple partitions, from which we shall then be able to read off the answer
for the general case.

The simplest case is that of the {\it homogeneous} Heisenberg
subalgebra,
which belongs to the partition $n=1+1+\cdots + 1$, when
$d=\lambda{d\over d\lambda}$ and the graded regular elements are
of the form
$$
\Lambda = \lambda^k {\rm diag}[y_1,y_2,\ldots,y_n]\,,
\qquad y_i\neq y_j\,,\quad\forall\,k\, .
\eqno(1.44)$$

The {\it principal} Heisenberg subalgebra belongs
to the other extreme case when $n$ is ``not partitioned
at all''.
In this case  $d=n\lambda {d\over d\lambda}+{\rm ad\,}{H_n}$
and the graded generators are the powers of the
``Drinfeld-Sokolov matrix'' $\La_n$.
We have
$$
\Lambda_n^{l+mn}=\lambda^m\Lambda_n^l\,,
\qquad
d(\Lambda_n^{l+mn})=(l+mn)\Lambda_n^{l+mn}\,,
\qquad
0\leq l\leq (n-1)\,,\ \ \forall\,m\,.
\eqno(1.45)$$
It is obvious that $\Lambda_n^{l+mn}$ is regular
if and only if $\Lambda_n^l$ is regular.
The Drinfeld-Sokolov matrix $\La_n$  itself is {\it regular}
since its eigenvalues are the $n$ {\it distinct} $n$th-roots of
$\lambda$, and from this one also easily verifies the following
by looking at the eigenvalues of $\Lambda_n^l$.

\medskip
\noindent
{\bf Lemma 1.4.}
{\it The element $\Lambda_n^l$ ($1\leq l \leq (n-1)$) is regular
if and only if $n$ and $l$ are relatively prime.}
\medskip

\medskip
Consider now a partition of the type
$$
n=n_1+n_2 \ , \qquad\hbox{with}\qquad n_1>n_2>1\ .
\eqno(1.46)$$

\noindent
{\bf Lemma 1.5.} {\it In the case (1.46) there is no graded regular
element in the Heisenberg subalgebra.}
\medskip

\noindent
{\it Proof.}
As candidates for graded regular elements, it is enough
to consider the matrices of the form
$$
\Lambda = \left[\matrix{ y_1\La_{n_1}^{l_1}&{}\cr
                {}&y_2\La_{n_2}^{l_2}\cr}\right]
\qquad\hbox{with}\qquad y_1 y_2\neq0\,,\quad l_i\neq 0 \bmod n_i\ .
\eqno(1.47)$$

\noindent
Case (i): Assume that $n_1$ and $n_2$ are relatively prime. We can
check from the definition of the grading that there is no graded
element of the form (1.47), since
$$
\left[\matrix{ \La_{n_1}^{l_1}&{}\cr
                {}&0_{n_2}\cr}\right]
\qquad\hbox{and}\qquad
\left[\matrix{ 0_{n_1}&{}\cr
                {}&\La_{n_2}^{l_2}\cr}\right]
\eqno(1.48)$$
have different grades.

\noindent Case (ii): If $n_1$ and $n_2$
are not relatively prime and $m>1$ is their greatest common divisor
then the graded elements of the form (1.47) are those for which
$$
l_1 = k{n_1\over m}
\qquad\hbox{and}\qquad
l_2=k{n_2\over m}\,,
\qquad\hbox{$k$: any integer}.
\eqno(1.49)$$
This implies by Lemma 1.4 that neither of the $\La_{n_i}^{l_i}$ ($i=1,2$)
is regular,  and therefore there is no  graded regular element
of the type (1.47) either. \quad  $\square$
\medskip
Let us also note the following rather obvious fact.

\noindent
{\bf Lemma 1.6.} {\it There is no graded regular element in the Heisenberg
subalgebra if the partition is of the type
$$
n=m+1+\cdots + 1 \,, \qquad 1<m<(n-1)\,,
\eqno(1.50)$$
i.e. if it  consists of a ``non-singlet'', $m$, and
more than one ``singlets'' ($1$'s).}

\medskip

It follows from the ``block structure'' of
the Heisenberg subalgebras given by (1.39)-(1.43) above  that graded regular
elements do not exist for any of those partitions which contain a
subset of the type appearing in Lemmas 1.5 and 1.6.
Hence the only cases not excluded are the ``partitions into equal
blocks'',
$$
n=p r = \overbrace{r+\cdots +r}^{p\;\rm times}\ , \qquad
\eqno(1.51)$$
and the cases ``equal blocks plus a singlet'',
$$
n=pr+1=\overbrace{r+\cdots +r}^{p\;\rm times}+1\,.
\eqno(1.52)$$
On the other hand, by inspecting the eigenvalues
of the generators of the Heisenberg subalgebras,
we can establish that graded regular elements do indeed
exist in the cases (1.51), (1.52), and
the following theorem gives the complete list.

\medskip
\noindent
{\bf Theorem 1.7.} {\it Graded regular elements exist only in those
Heisenberg subalgebras of $\ellg$ which belong to the special partitions
(1.51) or (1.52). In the equal block case (1.51)
with $r>1$ the graded regular elements are of the form
$$
\La=\lambda^m
\left[\matrix{ y_1 \La_{r}^{l}&{}&{}&{}\cr
              {}&y_2\La_{r}^{l}&{}&{}\cr
              {}&{}&\ddots&{}\cr
             {}&{}&{}&y_p\La^{l}_{r}\cr}\right] \, ,
\eqno(1.53)$$
where
$$
1\leq l\leq (r-1),\qquad
y_i\neq 0,\qquad y_i^r\neq y_j^r \qquad  i,j=1,\ldots,p\,,
\quad i\neq j\,,
$$
with $l$ relatively prime to $r$ and $m$ any integer.
The element $\Lambda$ is of grade $(l+mr)$ with respect to
the grading operator given by (1.42) where $N=r$ and}
$$
H={\rm diag}[\overbrace{H_r,H_r,\ldots,H_r}^{p\;\rm times}]\,.
\eqno(1.54)$$
{\it In the equal-blocks-plus-singlet case (1.52),
the graded regular elements are those
$n\times n$ matrices which
contain an $(n-1)\times (n-1)$ block of the form
given by (1.53) in the ``top-left corner''
and an arbitrary entry in the
``lower-right corner''.
The relevant grading operator is
given by (1.42) with $N=r$,
$$
H={\rm diag}[\overbrace{H_r,H_r,\ldots,H_r}^{p\;\rm times},0]
\qquad\qquad \hbox{\it if $r$ is odd}\,;
\eqno(1.55{\rm a})$$
and with $N=2r$,}
$$
H={\rm diag}[\overbrace{2H_r,2H_r,\ldots,2H_r}^{p\;\rm times},0]
\qquad\qquad\hbox{\it if $r$ is even}\,.
\eqno(1.55{\rm b})$$

\medskip
\noindent
{\it Remark 1.3.} Let us designate
the ordered eigenvector basis of the matrix $H$ in (1.54) as
$$
X_{j,1},X_{j-1,1},\ldots,X_{-j,1};
X_{j,2},X_{j-1,2},\ldots,X_{-j,2};
\ldots;
X_{j,p},X_{j-1,p},\ldots,X_{-j,p}.
\eqno(1.56)$$
Here $j=(r-1)/2$, the first index is the eigenvalue and the
second one orders the $r\times r$ blocks.
It is often convenient to use the re-ordered basis
$$
X_{j,1},X_{j,2},\ldots,X_{j,p};
X_{j-1,1},X_{j-1,2},\ldots,X_{j-1,p};
\ldots;
X_{-j,1},X_{-j,2},\ldots,X_{-j,p}.
\eqno(1.57)$$
When expressed in the new basis the matrix $H$ of (1.54)
may be written as $H_r\otimes 1_p$ and
the graded regular element $\Lambda$ given by (1.53) takes the form
$$
\lambda^m \Lambda_r^l\otimes D\,,
\qquad\hbox{where}\qquad
D={\rm diag\,}(y_1,y_2,\ldots,y_p)\,.
\eqno(1.58)$$
\medskip

In the following section we shall consider reductions based
on the grade $1$ regular elements belonging to the equal block
case.

\vfill\eject


\centerline{\bf 2. Equal block reduction of the AKS system }
\centerline{\bf to the matrix Gelfand-Dickey hierarchy}
\medskip

We have seen that graded regular elements exist only in
those Heisenberg subalgebras of $\ellg$ which correspond to
partitions into equal blocks or equal blocks plus a singlet.
The purpose of this section is to study in some detail
 symmetry reductions of the general
AKS system that are based upon
the grade $1$ regular elements of the
Heisenberg subalgebras belonging to the partitions into equal
blocks, $n=pr$, generalizing the $p=1$ case
considered in [1].
The final result of the analysis below is that the
reduction of the bihamiltonian manifold ${\cal M}:={\cal M}_{0,1}$
carrying the commuting family of ${\rm ad}^*$-invariant Hamiltonians yields
the $p\times p$ matrix  version of the well-known
Gelfand-Dickey r-KdV hierarchy.
More exactly, we shall establish the following:

\smallskip

\item{1.}
The reduced phase space is the space of
$r$th order, $p\times p$ matrix differential operators
carrying the first and second {\it Gelfand-Dickey
Poisson brackets}. The second Poisson bracket algebra
is an example of a classical ${\cal W}$-algebra.

\smallskip

\item{2.}
The commuting hierarchy of Hamiltonians resulting from the
monodromy invariants is given by the componentwise
residues of the fractional (including integral) powers of
the {\it pseudo-differential operators} obtained by
{\it diagonalizing} the matrix differential operators.

\smallskip

\noindent
The exact statements are given by Theorem
2.4 and Corollary 2.11 in Sections 2.2 and 2.3.
These results generalize the analogous results proven
by Drinfeld and Sokolov for the scalar case $p=1$.
We shall in fact use many of their methods, but
at the same time introduce some simplifications (at least to our taste)
in the proofs.

\medskip
\noindent
{\it Remark 2.1.}
For $p>1$ the subhierarchy provided by
the trace-residues of the {\it fractional} powers of
the matrix differential operators is not exhaustive,
since it does not include the Hamiltonians obtained from
the {\it integral} powers of the corresponding diagonal
pseudo-differential operators, which also appear in Corollary 2.11.

\bigskip
\noindent
{\bf 2.1. A local symmetry reduction of the AKS system}
\medskip

After reordering the basis as explained previously,
the generators of our Heisenberg subalgebra are the $n\times n$
matrices of the form $\Lambda_r^k \otimes D$,
where $\Lambda_r^k$ is the $k^{\rm th}$ power of the $r\times r$
DS matrix, $k$ is an arbitrary integer,
and $D$ is an arbitrary $p\times p$ diagonal matrix.
The generator $\Lambda_r^k\otimes D$ is of grade $k$ under the
grading defined by
$d= r \lambda{d\over d\lambda} + {\rm ad\,} H$ with
$$
H=H_r\otimes 1_p=
{\rm diag\ }[\,j 1_p\,,\, (j-1)1_p\,,\,\ldots \,,\,-(j-1)1_p\,,\, -j 1_p\,]\,,
\quad j={(r-1)\over 2}.
\eqno(2.1)$$

Choose a {\it grade $1$ regular element} of
the Heisenberg algebra,
(cf.~(1.53), (1.58));
i.e.,~an element
$$
\Lambda :=\Lambda_r\otimes D
\eqno(2.2)$$
such that $D^r$ has distinct, non-zero eigenvalues.
Define the $\lambda$-independent constant matrices
$C_0$ and $C_1$ by the equality
$$
\Lambda = \Lambda_r\otimes D := C_0 + \lambda C_1\ .
\eqno(2.3{\rm a})$$
More explicitly, these matrices are given
in block form by
$$
\Lambda
=\left[\matrix{
0&D&0&\cdots&0\cr
\vdots&0&D&\ddots&\vdots\cr
\vdots&{}&\ddots&\ddots&0\cr
0&{}&{}&\ddots&D\cr
\lambda D&0&\cdots&\cdots&0\cr}\right]
:=C_0+\lambda C_1\,.
\eqno(2.3{\rm b})$$
We start reducing the general AKS system
carried by $\ell({\widetilde {{gl}_n}}^\wedge)^*$ by
confining ourselves to the space
${\cal M}$ consisting of elements of the form
$$
{\cal L}=\partial +J+\lambda C_1\,,
\qquad
J : S^1\to gl_n\ .
\eqno(2.4)$$
This is an example of a Poisson subspace of the type
${\cal M}_{0,1}$ considered in Sec.~1.1,
 and
therefore it carries the two compatible Poisson brackets
given by (1.24a,b).

Consider the decomposition
$$
gl_n=gl_n^- + gl_n^0 + gl_n^+
\eqno(2.5)$$
induced by the eigenvalues of ${\rm ad\,} H$, eq.~(2.1),
where the summands are the subalgebras of block lower triangular,
block diagonal,
and block upper triangular matrices (with $p\times p$ blocks),
respectively.
On account of the relation
$$
[gl_n^-\,,\,C_1]=0\ ,
\eqno(2.6)$$
the group ${\cal N}$ of transformations
$$
e^f : {\cal L}\mapsto e^{f}{\cal L}e^{-f}\,,
\qquad\hbox{with}\qquad  f : S^1\to  gl_n^-\ ,
\eqno(2.7)$$
is a  symmetry group of the AKS system carried by ${\cal M}$;
i.e.,~these transformations preserve the two Poisson structures
and the monodromy invariants.
Next we define a
{\it symmetry reduction of the AKS system}
using
${\cal N}$ in such a way as  to ensure the {\it locality}
of the reduced system.
That is,  consider the following two step reduction process,
which is an obvious generalization of the one used in [1].
First, we restrict our system to the ``constrained manifold''
${\cal M}_c\subset {\cal M}$, defined as the set of
${\cal L}$'s of the following special form:
$$
{\cal L}=\partial + (q+C_0)+\lambda C_1 =\partial +q+\Lambda\,,
\qquad
q: S^1\to (gl_n^-+gl_n^0)\ .
\eqno(2.8)$$
Here  $C_0$ is the constant matrix given by (2.3), and
$q$ is required to be block lower triangular.
(Note that with respect to the second Poisson structure (1.24b),
this is just fixing a level set of the moment map
generating the Hamiltonian group action (2.7), and the reduction
procedure is essentially that of Marsden-Weinstein at the level
of a Poisson manifold rather than a symplectic one.
With respect to the first Poisson structure (1.24a),
the constrained quantities determining the form of ${\cal L}$
in (2.8) are all Casimirs and hence this just determines a
Poisson submanifold.)
Second, we factorize this constrained manifold by the
symmetry group ${\cal N}$, defining the reduced phase space
$$
{\cal M}_{\rm red}={\cal M}_c/{\cal N}\ .
\eqno(2.9)$$
To put it another way,  we factorize out the
``gauge transformations'' generated by ${\cal N}$ by declaring
that only the ${\cal N}$-invariant functions of ${\cal L}$
are physical.
The nice features of this reduction are that

\medskip

\noindent
\item{i)}
   The monodromy invariants of ${\cal L}\in {\cal M}_c$
   can be computed algebraically as asymptotic series in $\lambda$
   which depend on $q$ through integrals of local densities formed
    from its components and their derivatives.
\smallskip

\noindent
\item{ii)}
    The compatible Poisson structures on ${\cal M}$ induce compatible
    Poisson structures on ${\cal M}_{\rm red}$.

\smallskip

\noindent
\item{iii)}
     The  gauge orbits in ${\cal M}_c$  allow for global, differential
     polynomial gauge sections, which give rise to complete sets of
     gauge invariant differential polynomials.

\medskip

Statement i) follows immediately
since we have chosen ${\cal M}_c$ so that the conditions
of Proposition 1.2 are satisfied.
Statement ii) means that the compatible Poisson brackets carried
by ${\cal M}$ can be consistently restricted to the {\it gauge invariant}
functions in $C^\infty({\cal M}_c)$,
whose space can be naturally identified with
$C^\infty({\cal M}_{\rm red})$.
This can be seen as a consequence of the Dirac theory of
reduction by constraints as follows.
We first note that, by choosing some basis $\{\gamma_i\}$ of $gl_n^-$, the
constraints defining ${\cal M}_c\subset {\cal M}$ can be written as
$$
\chi_i(x)=0\,
\qquad\hbox{where}\qquad \chi_i(x)= {\rm tr\,}\gamma_i (J(x)-C_0)\ .
\eqno(2.10{\rm a})$$
It  is easy to verify  that these constraints are {\it first class}; i.e.,
$$
\{ \chi_i (x)\,,\,\chi_k(y)\}{\vert_{{\cal M}_c}}=0\ ,
\eqno(2.10{\rm b})$$
for any of the compatible Poisson brackets on ${\cal M}$.
We next notice that the functions $\chi_i(x)$ are
the generating densities
(i.e., components of the moment map)
of the ${\cal N}$ symmetry transformations
with respect to the second Poisson bracket (1.24b).
Thus the theory of reduction by constraints
(which in this case is just the Poisson version
of Marsden-Weinstein reduction)
 tells us
to factorize the constrained manifold by these transformations;
the second Poisson bracket algebra closes on the gauge invariant functions
on ${\cal M}_c$,  inducing a Poisson structure on the factor space.
On the other hand, the $\chi_i$ do not generate
any transformations on ${\cal M}$  under the first Poisson bracket (1.24a);
i.e.,~they are Casimir functions.
Therefore the first Poisson bracket can in principle already be restricted to
$C^\infty ({\cal M}_c)$ without any factorization by ${\cal N}$.
Then ${\cal N}$ becomes a group of Poisson  maps with respect to the
restricted bracket, which can further be reduced to a Poisson bracket
on the invariant functions.
In this way, we naturally obtain two induced Poisson brackets
on ${\cal M}_{\rm red}$ from those on ${\cal M}$,
and the induced Poisson brackets are compatible because the
original brackets (1.24a,b) were compatible.

One is always interested in gauge invariant objects and convenient
gauge fixings when describing systems with gauge symmetries.
In the present example, as in the $p=1$ case of [1],
a convenient gauge section is defined by the subspace
$$
V:=\{\, \partial + \sum_{i=1}^r e_{r,i}\otimes v_{r-i+1}+\Lambda\,\,
\vert\, v_k : S^1\to gl_p\,\}\,,
\eqno(2.11)$$
i.e., the space of ${\cal L}$'s
in which the matrix function $q$ is allowed to have non-zero
($p\times p$ block) entries only in the last row.
This global section of the gauge orbits can be reached
 from an arbitrary point ${\cal L}=(\partial +q+\Lambda ) \in {\cal M}_c$ by a
unique gauge transformation  that depends on $q$ in a differential
polynomial way.
It follows that the components of the $v_i$ provide
 a basis (generating set) for the
gauge invariant differential polynomials which can be
formed from the components of $q$.

\medskip
\noindent
{\it Remark 2.2.}
A distinguished gauge invariant differential polynomial
is obtained by restricting
$$
{\cal V}_H = {1\over 2} {\rm tr} (J^2) + {\rm tr\,}(H J^\prime )
\eqno(2.12)$$
 from ${\cal M}$ to ${\cal M}_c$.
The density ${\cal V}_H$ satisfies the Virasoro algebra under
the second Poisson bracket, and therefore it generates an action of
${\rm Diff}(S^1)$ on ${\cal M}$ that survives the present reduction.
Since it contains this Virasoro density,
the second Poisson bracket algebra of the gauge invariant differential
polynomials can be regarded as an extended conformal algebra;
that is, a classical ${\cal W}$-algebra.
This ${\cal W}$-algebra is a member of a
natural family of extended conformal algebras that recently has been
studied in [4], [5].
The reader is referred to [5] for
a detailed description of differential polynomial
gauge fixings, like the gauge (2.11),  and for the
related construction of a generating set for the
${\cal W}$-algebra  consisting of ${\cal V}_H$ and conformal tensors.
\medskip

In the next section we shall also need the
``block diagonal gauge'' given by
$$
\Theta := \{\,\partial +\sum_{i=1}^r e_{i,i}\otimes
\theta_i +\Lambda\,\,\vert\, \theta_i : S^1\to gl_p\,\}\, .
\eqno(2.13)$$
It should be noted that
$\Theta$ defines only a partial gauge fixing,
which has a finite dimensional residual gauge freedom.
Of course it is nevertheless true that any gauge invariant element of
$C^\infty({\cal M}_c)$ can be recovered from its restriction to $\Theta$.
In particular,  the Poisson bracket of any two gauge invariant functions in
$C^\infty({\cal M}_c)$ can be recovered from its restriction to $\Theta$,
and the nice feature is that for the second Poisson bracket this restriction
can be computed in terms of the ``free current algebra'' of
the $\theta_i$'s.
For later reference we summarize this fact as a lemma.
The proof is similar to that for the $p=1$ case [1].

\medskip\noindent
${\bf Lemma\  2.1.}$
{\it Let $\varphi,\psi\in C^\infty ({\cal M}_c)$ be gauge invariant functions
and consider $\xi = \{\varphi,\psi\}_2\in C^\infty({\cal M}_c)$,
which is well defined and is also gauge invariant.
Let us denote the restriction of these functions to
$\Theta$ by $\bar \varphi, \bar \psi$ and $\bar \xi$, respectively.
Then we have}
$$
\bar \xi (\th_1,\ldots,\th_r)=
\sum_{i=1}^r \int_{S^1} {\rm tr\,}\left(\theta_i
[{\delta \bar\varphi\over \delta \theta_i} ,
{\delta \bar\psi\over \delta \theta_i}] +
\left({\delta \bar\varphi\over \delta \theta_i}\right)^\prime
{\delta \bar\psi\over \delta \theta_i}\right)\ .
\eqno(2.14)$$
\medskip
\noindent

We consider the decomposition $J=J_-+J_0+J_+$
defined by means of (2.5) and write out formula (1.24b)
in terms of the partial functional derivatives
corresponding to these variables.
We then compute the value of $\{\varphi,\psi\}_2$ at
an arbitrary point on the ``gauge slice'' $\Theta$
of the block diagonal gauge, where
$J_-=0$, $J_0={\rm diag\,}[\theta_1,\ldots,\theta_r]$, $J_+=C_0$.
(For this computation we can extend $\varphi$ and $\psi$ from
${\cal M}_c$ to ${\cal M}$ in an arbitrary way since
they are invariant under the
gauge transformations generated by the first class
constraints specifying this constrained manifold.)
Using the grading structure, we find that
$$\eqalign{
{\bar \xi}(J_0):=
\{\varphi,\psi\}_2(J_0+C_0)
&=\int_{S^1} {\rm tr\,}\left(J_0
[{\delta \varphi\over \delta J_0} ,
{\delta \psi\over \delta J_0}] +
\left({\delta \varphi\over \delta J_0}\right)^\prime
{\delta \psi\over \delta J_0}\right)\cr
&=\sum_{i=1}^r \int_{S^1} {\rm tr\,}\left(\theta_i
[{\delta \varphi\over \delta \theta_i} ,
{\delta \psi\over \delta \theta_i}] +
\left({\delta \varphi\over \delta \theta_i}\right)^\prime
{\delta \psi\over \delta \theta_i}\right)\,,\cr}
\eqno(2.15)$$
where
${\delta \varphi\over \delta \theta_i}$,
${\delta \psi\over \delta \theta_i}$
are understood as the $p\times p$ blocks constituting
the block diagonal ${\delta \varphi\over \delta J_0}$,
${\delta \psi\over \delta J_0}$.
This immediately gives (2.14) since at the point
$J=J_0+C_0$ we have
${\delta \varphi\over \delta \theta_i}=
{\delta \bar \varphi\over \delta \theta_i}$,
and similarly for $\psi$.
\quad $\square$
\medskip

Now let $M$ be the space of $p\times p$
``matrix Lax operators'' of the following form:
$$
L=(-D^{-1})^r \partial^r +\sum_{i=1}^r u_i \partial^{r-i}\,,
\qquad  u_i : S^1\to gl_p \ .
\eqno(2.16)$$
As in the $p=1$ case [1], $M$ can serve as a natural model of the
reduced space ${\cal M}_c/{\cal N}$.
To see this let us consider the linear problem
$$
{\cal L}\Psi = {\cal L}
\pmatrix{\psi_1\cr
         \psi_2\cr
         \vdots\cr
         \psi_r\cr}=0\,,
\eqno(2.17)$$
for ${\cal L}$ given by (2.8), where
the entries $\psi_i$ of $\Psi$ are $p$-component column vectors.
This system of equations is covariant under (2.7) if we let
$\Psi$ transform as
$$
\Psi \rightarrow e^f \Psi \ .
\eqno(2.18)$$

Denoting
the $ij$th $p\times p$ block of $q$ by $q_{ij}$,
the system (2.17) can be
recast in the following form:
$$\eqalign{
L\psi_1 =&\lambda \psi_1\cr
\psi_2 =&(-D)^{-1} (\partial + q_{11}) \psi_1\cr
\psi_3=&(-D)^{-1} [q_{21} + (\partial +
q_{22})(-D)^{-1}(\partial +q_{11})]\psi_1\,,\cr
       {\phantom{=}}&\vdots \cr}
\eqno(2.19)$$
where $L$ is an operator of the form (2.16) whose coefficients
$u_i$ are uniquely determined differential polynomials in $q$.
Notice now that the component $\psi_1$ is invariant
under (2.18), because $f$ is a strictly block lower
triangular matrix.
This implies that the
potentials $u_i=u_i[q]$ entering in $L\psi_1 = \lambda \psi_1$
must also be invariant functions of $q$ under (2.18).
In other words, the operator $L$ attached to ${\cal L}$
by the equivalence of (2.17) and (2.19)
actually depends only on the ${\cal N}$-orbit of ${\cal L}$ in ${\cal M}_c$.
Thus it gives rise to a map  $m : {\cal M}_{\rm red} \rightarrow M$.
It is easy to see that this is a one-to-one map.
The inverse $m^{-1}$ can be given by attaching to an arbitrary operator
$L$ in (2.16) the unique orbit in ${\cal M}_c$ which
intersects the gauge section $V$ (2.11) in the point
$$
{\cal L} = \partial + \sum_{i=1}^r e_{r,i}\otimes v_{r-i+1}+\Lambda
\qquad\hbox{where}\qquad
v_i = (-1)^{r+1-i} D u_i D^{r-i}\ .
\eqno(2.20)$$
This equation provides an identification of the
space $M$ with the
gauge section $V$ and hence with the
reduced phase space ${\cal M}_{\rm red}$,
$$
M\simeq V\simeq {\cal M}_{\rm red}\ .
\eqno(2.21)$$

It follows from the above that the natural mapping
$\pi :\Theta\to M$ is given by the factorization formula:
$$
L=(-D^{-1})^r\pa^r+u_1\pa^{r-1}+\cdots + u_r=
(-D^{-1}(\pa+\th_r))\cdots (-D^{-1}(\pa+\th_1))\ ,
\eqno(2.22)
$$
generalizing the scalar case ([23], [1]).
Lemma 2.1 can be reformulated as saying
that $\pi$, which is the usual Miura map when $p=1$,
is a {\it Poisson mapping} if the space $\Theta$ carries the
Poisson bracket appearing on the right hand side
of (2.14) and $M$ carries the Poisson bracket induced
 from the second Poisson bracket on ${\cal M}$ via
the reduction and the identification $M\simeq {\cal M}_{\rm red}$.

\bigskip
\noindent
{\bf 2.2. The Gelfand-Dickey form of the reduced Poisson structures}
\medskip

In this section we prove a theorem
that establishes the equivalence of
the Poisson structures naturally
carried by the spaces ${\cal M}_{\rm red}$ and $M$.
This theorem will rely on the preliminary
Lemma 2.2 which follows\footnote*{
This lemma was explained to us by A. G. Reyman.}.
This concerns the Poisson
Lie group  property of the so-called
Sklyanin bracket as pointed out by Semenov-Tian-Shansky [20].

Let ${\cal A}$ be an associative algebra.
Suppose that $\Tr:{\cal A}\to {\bf C}$ is a
non-degenerate trace-form on ${\cal A}$;
i.e.,~a linear mapping with the property that the formula $\la a,b\ra=\Tr ab$
defines a non-degenerate, symmetric bilinear form on ${\cal A}$.
As usual, identify the space ${\cal A}$ with its dual ${\cal A}^*$
(or a subspace thereof) by means of the pairing defined by $\Tr$.
Suppose that $A$ and $B$ are disjoint  subalgebras of ${\cal A}$ such that
${\cal A}=A+B$ and that with respect to the pairing $\la\ ,\ \ra$
we have $A^\perp =A$ and $B^\perp =B$.
Let $P_A$ and $P_B$ be the projection maps on ${\cal A}$
defined by the splitting ${\cal A}=A+B$.
Since $A$ and $B$
turn into Lie subalgebras of ${\cal A}$ with respect to
the natural Lie algebra structure on ${\cal A}$ given by
$[a,b]=ab-ba$, then $R=P_A-P_B$ is a skew-symmetric r-matrix on ${\cal A}$,
and the AKS construction applies here too.

For any function $\varphi\in C^\infty({\cal A})$, we define
$\nabla_a \varphi$, its gradient at the point
$a \in {\cal A}$, by
$$
{d\over dt} \varphi(a+t b){\vert_{t=0}} =\la b,\nabla_a \varphi\ra =
 \Tr b \nabla_a \varphi\,,
\qquad \forall\,b\in {\cal A}\ .
\eqno(2.23)$$
In addition to the R Lie-Poisson bracket given by
$$
\{\varphi , \psi\}^{(1)}(a):=\Tr\left(a [X,Y]_R\right)\,,
\quad\hbox{where}\quad X=\nabla_a\varphi ,\,  Y=\nabla_a\psi\,,
\eqno(2.24)$$
we can also define a second Poisson bracket on ${\cal A}$:
$$
\{\varphi, \psi\}^{(2)}(a):={1\over 2}\Tr\left(YaR(Xa)-aYR(aX)\right)\,.
\eqno(2.25)$$
This Poisson bracket is known as the quadratic r-bracket,
the Sklyanin bracket, or the second Gelfand-Dickey bracket.
The proof of the Jacobi identity for the Poisson bracket (2.25) is easy if
we remember that because of the derivation property it
is sufficient to check it for
linear functions; i.e., functions of the form $f_{_X}(a)=\Tr(aX)$, $X\in {\cal
A}$.
In the proof we also use the fact that $R$ satisfies
the modified Yang-Baxter equation,
$$
[RX,RY] = R([RX,Y] + [X,RY]) - [X,Y] \qquad \forall\, X ,\, Y \in {\cal A}\,.
\eqno (2.26)$$
Note that the $R$ Lie-Poisson bracket and the
quadratic r-bracket are {\it compatible}.

\medskip
\noindent
{\bf Lemma 2.2.}
{\it With respect to the quadratic r-bracket on ${\cal A}$, the mapping
$m:{\cal A}\times {\cal A}\to {\cal A}$ given by $m(a,b)=ab$
is a Poisson mapping
when ${\cal A}\times {\cal A}$ has the product Poisson structure corresponding
to the quadratic r-bracket on each of its components.}
\medskip
\noindent
{\it Proof.} For a pair of elements $X,Y\in {\cal A}$,
consider the linear functions on ${\cal A}$ given by
$$
f_{_X}(l)=\Tr lX \ ,\ f_{_Y}(l)=\Tr lY \,.
$$
Then
$$
m^*f_{_X}(a,b)=\Tr abX
\qquad\hbox{and}\qquad m^*f_{_Y}(a,b)=\Tr abY \ .
$$
Variation of $m^*f_{_X}$ with respect to $a$ for fixed $b$ gives
$$
{\nabla_a m^*f_{_X}}{\vert_{(a,b)}}=bX\ ,
$$
similarly
$$
{\nabla_b m^*f_{_X}}{\vert_{(a,b)}}=Xa\ .
$$
Thus
$$\eqalign{
\{m^*f_{_X} , m^*f_{_Y}\}^{(2)}_{{\cal A}\times {\cal A}}(a,b)=&
{1\over 2}\Tr\left(bYaR(bXa)-abYR(abX)\right)\cr
+&{1\over 2}\Tr\left(YabR(Xab)-bYaR(bXa)\right)\cr
=&{1\over 2}\Tr\left(YabR(Xab)-abYR(abX)\right)
=\left(m^*\{f_{_X},f_{_Y}\}^{(2)}_{\cal A}\right)(a,b)\;.\cr}
\eqno{\square}
$$
\medskip

We shall apply
this lemma to the following  example:
$$
{\cal A} = \{\,X = \sum_{s=-\infty}^N X_s\partial^s\,
\vert\, \forall\, X_s : S^1 \to p\times p \ {\rm matrices},
\ \forall\,N\ge 0\,\}:=  GD_M,
\eqno (2.27)
$$
the space of pseudo-differential operators with
$p\times p$ matrix coefficients.
We call this space
$GD_M$ for ``matrix Gelfand-Dickey''.
\noindent
Multiplication of matrix pseudo-differential operators is defined
in the usual way ([14], [16]).
\noindent
The trace form, $\Tr: GD_M \to {\bf C}$, is given by
$$
\Tr X :=  \int_{S^1} \, \tr\,{\rm res}(X) =  \int_{S^1} \, \tr\, X_{-1}\,,
\eqno (2.28)$$
where $\tr$ is the ordinary matrix trace.
The two subalgebras $A$ and $B$ in this case are given by
$$
A := GD_{M_+} = \{\,X = \sum_{s=0}^N X_s\partial^s\,\}\,,
\qquad
B := GD_{M_-} = \{\,X = \sum_{s=-\infty}^{-1} X_s\partial^s\,\}
\eqno (2.29)
$$
It is usual to write
$P_A(X) = X_+$ and $P_B(X) = X_-$.
One sees by inspection that for any $k>0$, the space
$$
M_{0,k} := \{\, X = X_0\partial^k + X_1\partial^{k-1} + \cdots + X_k\,
\vert\, X_0 \ {\rm fixed}\,\} \subset GD_M
\eqno(2.30)$$
is a Poisson subspace with respect to both
Poisson brackets (2.24) and (2.25).

In particular, the space $M$ of operators $L$ defined
by (2.16) is such a subspace, on which the two Poisson
brackets take the form
$$
\eqalignno{
\{\varphi,\psi\}^{(1)}(L)&=
\int_{S^1}\tr{\rm res\,}\left(L [Y_-,X_-]\right)\,,
&(2.31)\cr
\{\varphi,\psi\}^{(2)}(L)&=
\int_{S^1}\tr{\rm res\,}\left(YL(XL)_+ -LY(LX)_+\right)\,,
&(2.32)\cr}
$$
where $X:=\nabla_L\varphi$, $Y:=\nabla_L\psi$
for $\varphi$, $\psi\in C^\infty(M)$.
It is now easy to prove the following
corollary of Lemma 2.2.

\medskip
\noindent
{\bf Corollary 2.3.}
{\it The mapping
$\pi: \Theta \to M$ given by eq.~(2.22) is a
Poisson mapping where the Poisson bracket of
$f,h\in C^\infty(\Theta)$ is given by
$$
\{f,h\}(\th_1,\ldots,\th_r)=\sum_{i=1}^r \int_{S^1} \tr \left(
\th_i [{\delta f\over \delta \th_i},{\delta h\over \delta \th_i}] +
\left({\delta f\over \delta \th_i}\right)^\prime
{\delta h\over \delta \th_i}\right)\,,
\eqno(2.33)
$$
and the Poisson bracket
of $\varphi,\psi\in C^\infty(M)$ is given by the second
Gelfand-Dickey bracket (2.32) on the manifold $M$.}

\medskip
\noindent
{\it Proof.}
It is sufficient to check that on the Poisson
subspace  $\{-D^{-1}(\pa +\theta)\}$
of $GD_M$ (a subspace of type $M_{0,1}$)
the Poisson bracket (2.25) becomes just the $r=1$
case of (2.33), that is,
if $\varphi,\psi\in C^\infty(\{\theta\})$, then
$$
\{\varphi,\ \psi\}^{(2)}(-D^{-1}(\pa +\theta))=
\int_{S^1} \tr \left(\theta
[{\delta \varphi\over \delta \theta},{\delta \psi\over \delta \theta}]
+\left({\delta \varphi \over \delta \theta}\right)^\prime
{\delta \psi\over \delta \theta}\right)\,\ .  \qquad \square
\eqno(2.34)
$$
\medskip

We may now state the main result of this section.

\medskip
\noindent
{\bf Theorem\  2.4.}
{\it Under the identification ${\cal M}_{\rm red}\simeq M$ given
in (2.21), the Poisson brackets on the space
${\cal M}_{\rm red}={\cal M}_c/{\cal N}$
obtained by reduction from the
Poisson brackets (1.24a) and (1.24b)
on ${\cal M}$ are equal respectively to the first and second
Gelfand-Dickey Poisson brackets
given by (2.31) and (2.32).}
\medskip

\noindent
{\it Proof.}
Consider the one parameter group of transformations
on ${\cal M}$ defined by
$$
g_\tau : {\cal L}\mapsto ({\cal L}-\tau C_1)\, , \qquad \tau \in {\bf R}
\eqno(2.35)$$
These transformations preserve ${\cal M}_c\subset {\cal M}$
and commute with the action of ${\cal N}$.
Thus we have a corresponding one parameter
group of transformations $\{\bar g_\tau\}$ on $M\simeq{\cal M}_c/{\cal N}$,
 operating as
$$
\bar g_\tau : (L\mapsto L+\tau 1_p)
\qquad (1_p : \hbox{the $p\times p$ unit matrix})\,.
\eqno(2.36)$$
We now recall that $\{\ ,\ \}_1$
given by (1.24a) is the Lie derivative of $\{\ ,\ \}_2$
given by (1.24b) with respect to the vector field
generating the flow (2.35).
It follows that a similar relation holds for the corresponding
induced brackets on $M\simeq{\cal M}_{\rm red}$ with respect to
the generator of the projected flow (2.36).
On the other hand, one also sees by inspection that
the first Gelfand-Dickey bracket on $M$;
i.e.,~the restriction of (2.24) given by (2.31),
is the Lie derivative of the second Gelfand-Dickey
bracket (2.32) with respect to the generator of (2.36).
Therefore the equality of the ``first brackets''
claimed in the theorem
follows if we prove the equality of the ``second brackets''.
This latter is obtained as an immediate consequence of
Lemma 2.1 of and the above corollary to Lemma 2.2. \quad  $\square$

\medskip
\noindent
{\it Remark 2.3.}
It is clear from the proof that Theorem 2.4 remains valid
if we let $D$ be any invertible $p\times p$
matrix in the construction.
The choice of a graded regular element for $\Lambda$ in (2.2)
is made in order to guarantee the existence of local monodromy
invariants by Proposition 1.2.
In the next section this assumption will
 be fully used in determining the Hamiltonians of the
reduced AKS system.

\medskip
\noindent
{\it Remark\ 2.4.}
Let $R_c:GD_M\to GD_M$ be the mapping defined by right-multiplication
by the invertible (constant) matrix $c$; i.e.,
$R_c(a)=ac$ for any $a\in GD_M$.
This is a one-to-one mapping
that preserves the second Gelfand-Dickey
bracket and carries the first Gelfand-Dickey bracket into the
Poisson bracket $\{\ ,\ \}_c^{(1)}$ on $GD_M$ given by
$$
\{\varphi\,,\,\psi\}^{(1)}_c(a)=\Tr(ac^{-1}[cX,cY]_R)\,,
\quad\hbox{where}\quad
X=\nabla_a\varphi,\ Y=\nabla_a\psi\,,
\eqno(2.37)$$
which is compatible with $\{\ ,\ \}^{(2)}$ for any $c$.
By means of this mapping, with $c= (-D)^r$, we can map the
space $M$ of $L$'s in eq.~(2.16) onto
the space $\widetilde M$ consisting of the $r$th order
differential operators  of the form
$$
{\widetilde L}=\pa^r +\sum_{i=1}^r{\tilde u}_i \pa^{r-i}\,,
\qquad  \tilde u_i: S^1\to gl_p\,.
\eqno(2.38)$$
This is the reduced space we would have
obtained by replacing the matrix $D$ by the unit matrix in the
reduction procedure described in the previous section.
(The results of Theorem 2.10 and Corollary 2.11 below,
which characterize the reduced monodromy invariants,
would not then be applicable.)
The one-to-one mapping between $M$ and $\widetilde M$
provided by $R_c$
can be made into an isomorphism of bihamiltonian manifolds
if one lets $M\subset GD_M$ carry the restrictions of the two Gelfand-Dickey
brackets, while ${\widetilde M}\subset GD_M$ carries the
restriction of the ``modified first bracket'' $\{\ ,\ \}_c^{(1)}$
with $c=(-D)^r$, plus the restriction of the
second Gelfand-Dickey bracket.

\bigskip
\noindent
{\bf 2.3. Computation of the Hamiltonians}
\medskip

In the previous section, we identified the Poisson
brackets carried by the reduced space
${\cal M}_{\rm red}\simeq M$ as the first and second Gelfand-Dickey
Poisson brackets.
In this section we establish a description of the Hamiltonians of the
reduced AKS system in terms of the reduced-space
variables $u_1,\ldots,u_r$.
This will result from a computation of the
eigenvalues of the monodromy matrix of the
linear problem ${\cal L}\Psi = 0$ for an
arbitrary ${\cal L}\in V$; i.e., for
$$
{\cal L} = \partial + C_{0} +\lambda C_{1} +
\pmatrix{0&\cdots&0\cr
         \vdots&{}&\vdots\cr
         0&\cdots&0\cr
         v_{r}&\cdots&v_{1}\cr}.
\eqno(2.39)$$

Define the matrix $\Delta$ by
$$
\Delta:=(-D)^{-1}\,.
\eqno(2.40{\rm a})$$
Note that $\Delta^r$ is a nondegenerate and
invertible diagonal matrix, by eq.~(2.2).
Let
$$
L=\Delta^r\pa^r+u_1\pa^{r-1}+\cdots+u_r
\eqno(2.40{\rm b})$$
be the element of $M$ corresponding to ${\cal L}\in V$.
If
$\{\phi_{\alpha}\}_{\alpha = 1, \ldots,rp}$ is
a complete set of independent ($p$-component column vector)
solutions to
$$
L\phi = \lambda \phi\,,
\eqno(2.41)$$
define the $n\times n$ matrix $\Phi$ by
$$
\Phi :=
\pmatrix{\phi_{1}&\cdots&\phi_{rp}\cr
        \Delta\phi'_{1}&\cdots&\Delta\phi'_{rp}\cr
        \vdots&{}&\vdots\cr
        \Delta^{r-1}\phi^{(r-1)}_{1}&\cdots&\Delta^{r-1}\phi^{(r-1)}_{rp}\cr}.
\eqno(2.42)$$
Then the columns of $\Phi$
are a complete set of solutions to ${\cal L}\Psi = 0$ and
$T=\Phi(2\pi)\Phi(0)^{-1}$ is the monodromy matrix,
whose invariants will be computed.

Since $\Delta^r$ is nondegenerate and diagonal, we can,
through the usual, recursive approach,
 find an operator $g$ of the form
$$
g = I + \sum_{i=1}^\infty g_i\partial^{-i}\,,
\eqno(2.43)$$
with $g_i(x+2\pi) = g_i(x)$ for all $i$, such that
$$
L = g\hat Lg^{-1}
\eqno(2.44)$$
with $\hat L$ a {\it diagonal} matrix pseudo-differential operator;
i.e.,~with $\hat L$ of the form
$$
\hat L = \Delta^r\partial^r + \sum_{i=1}^{\infty}a_{i}\partial^{r-i}\,,
\qquad a_i:\hbox{all diagonal matrices}.
\eqno(2.45)$$
For example, if we require
the $g_i$'s to be off-diagonal
then we can recursively uniquely determine
both the $g_i$'s and the $a_i$'s as differential
polynomials in the $u_i$'s, by comparing the two sides
of $Lg=g\hat L$ term-by-term according to powers of $\partial$.
Later we shall need that
$$
a_1=[u_1]_{\rm diag}\,.
\eqno(2.46)$$

Fix $-\zeta$ to be any $r$th root of $\lambda$,
$$
(-\zeta)^r = \lambda\,.
\eqno(2.47)$$
Consider the $p\times p$ diagonal matrix
asymptotic series
$$
\hat\psi(x,\,\zeta) = e^{\zeta Dx}
(\hat\phi_0(x) + \zeta^{-1}\hat\phi_1(x) + \cdots)
\qquad \hbox{for} \quad  \zeta\sim\infty
\eqno(2.48)$$
satisfying the equation
$$
(\hat L\hat\psi)(x,\zeta)=(-\zeta)^r\hat\psi(x,\zeta)
\eqno(2.49)$$
in the following sense (see e.g.~[24]).
Using the definition
$$
(\pa^{-s} e^{\zeta Dx}):=(\zeta D)^{-s}e^{\zeta Dx}  \eqno(2.50)
$$
for $s$ any integer,
and extending this in the obvious way to
pseudo-differential operators,
we can write
$$
\hat\psi(x,\,\zeta) = ({\cal D}e^{\zeta Dx})\,,
\eqno(2.51{\rm a})
$$
where
$$
{\cal D} =d_0(x) +  d_1(x)\pa^{-1} + \cdots
\eqno(2.51{\rm b})$$
is a uniquely determined diagonal pseudo-differential operator.
Then the action of any
pseudo-differential operator
$P\in GD_M$ on $\hat \psi$ is defined
by multiplication of pseudo-differential operators,
$$
(P \hat\psi)(x,\,\zeta) :=
((P {\cal D})e^{\zeta Dx})\,.
\eqno(2.52)$$
The left hand side of equation (2.49) is understood in this sense.
If we assume that $\det\hat\phi_0(x)\neq 0$ in (2.48),
then $\hat\psi$ is uniquely determined up to
multiplication by an $x$-independent diagonal
matrix of the form
$$
c(\zeta)=c_0+\zeta^{-1}c_1+\cdots\,,\
\qquad\hbox{with}\qquad  \det c_0\neq 0\,.
\eqno(2.53)$$

At this point we make use of a result given
in [1] (Theorem 2.9, pages 1986-7).
The proposition below states the result in
terms of the diagonal
pseudo-differential operator $\hat L$, and therefore we
are dealing with
$p$ different copies of the scalar case.

\medskip
\noindent
{\bf Proposition 2.5.} {\it If $\hat L$ is of the form (2.45),
 $\partial$ may be expressed as:}
$$
\partial = -D \LPR +
\sum_{i=0}^{\infty} F_i (-D \LPR)^{-i}\,,
\qquad (-D=\Delta^{-1})\,,
\eqno(2.54)
$$
{\it with
$$
F_0 = -{1\over r}(-D)^r a_{1}
\quad\hbox{\it and}\quad
\int_0^{2\pi}
\left(kF_k +(-D)^k \res ({\hat L}_1^{k/r}) \right) = 0\,,
\quad\hbox{\it for $k>0$},
\eqno(2.55)
$$
where $\hat L_1^{1/r}$ is the unique diagonal
pseudo-differential operator such that}
$$
{\hat L}_1^{1/r}=\Delta \pa +\sum_{i=0}^\infty b_i \pa^{-i}\,
\qquad\hbox{and}\qquad ({\hat L}_1^{1/r})^r=\hat L\,.
\eqno(2.56)$$
\medskip

A further argument from [1] can be applied directly to deduce from (2.49)

\medskip
\noindent
{\bf Lemma 2.6.}
$$
(\LPR \hat\psi)(x,\zeta) = -\zeta \hat\psi(x,\zeta).
\eqno(2.57)
$$
\medskip

Applying $\partial$ to $\hat\psi$ and making use of
Proposition 2.5 and Lemma 2.6,
we get
$$
\hat\psi^\prime =\left( \zeta D +
\sum_{i=0}^\infty F_i(\zeta D)^{-i}\right)\hat\psi\,.
\eqno(2.58)$$
This equation can be solved, giving
$$
\hat\psi(x,\zeta) =
\exp\left(
\zeta D x + \sum_{k=0}^\infty (\zeta D)^{-k} \int_0^x F_k \right)
\hat\psi(0,\zeta)\,.
\eqno(2.59)$$
It follows that
$$
\hat\psi(x+2\pi,\zeta)=\hat\psi(x,\zeta)\gamma(\zeta)\,,
\eqno(2.60)$$
where
$$
\gamma(\zeta)=
\hat\psi(0,\zeta)^{-1}
\exp\left(
2\pi\zeta D  + \sum_{k=0}^\infty (\zeta D)^{-k} \int_0^{2\pi} F_k \right)
\hat\psi(0,\zeta)\,.
\eqno(2.61)$$

Define the $p\times p$ matrix asymptotic series
$\psi(x,\zeta)$ by
$$
\psi(x,\zeta):=(g\hat\psi)(x,\zeta)\,,
\eqno(2.62)$$
where $g$ is the pseudo-differential operator appearing in (2.43).
We then have the self-evident

\medskip
\noindent
{\bf Lemma 2.7.}
{\it The columns of $\psi$ defined by (2.62) are solutions of (2.41).}
\medskip

If we expand $g$ in descending powers of
$-\LPR $
instead of descending powers of $\pa$ we have
$$
g= {\rm I} + m_1(-\LPR)^{-1} + m_2 (-\LPR)^{-2} + \cdots,
\eqno(2.63)$$
for some $m_1,m_2,\ldots$, with $m_i(x+2\pi) = m_i(x)$ for all $i$.
Then (2.57) gives
$$
\psi(x,\zeta)=(g\hat\psi)(x,\zeta)=
m(x,\zeta)\hat\psi(x,\zeta)\,,
\eqno(2.64{\rm a})$$
where
$$
m(x,\zeta):=(I+m_1(x)\zeta^{-1}+m_2(x)\zeta^{-2}+\cdots).
\eqno(2.64{\rm b})$$
  From this we obtain

\medskip
\noindent
{\bf Lemma 2.8.}
$$
\psi(x+2\pi,\zeta)=\psi(x,\zeta)\gamma(\zeta)\,.
\eqno(2.65)$$
\medskip

\smallskip
Now consider the set of $r$ independent roots of $\lambda$,
given by
$$
\{\,-\zeta_k = -e^{2\pi i k/r}\zeta\, \}_{k=0,\ldots, r-1}\,,
\eqno(2.66)$$
and define $\psi_i$ by
$$
\psi_i(x,\zeta):=\psi(x,\zeta_i)=m(x,\zeta_i)\hat\psi(x,\zeta_i)
\qquad i=0,1,\dots,r-1.
\eqno(2.67)$$
The point of this construction is:

\medskip
\noindent
{\bf Lemma 2.9.}

{\it The columns of the $n\times n$ matrix
$$
\Phi =   \pmatrix{\psi_0&\cdots&\psi_{r-1}\cr
                  \Delta\psi'_0&\cdots&\Delta\psi'_{r-1}\cr
                   \vdots &{}&\vdots\cr
 \Delta^{r-1}\psi^{(r-1)}_0&\cdots&\Delta^{r-1}\psi^{(r-1)}_{r-1}\cr}
\eqno(2.68)$$
are a complete set of solutions to ${\cal L}\Psi=0$.}
\medskip
\noindent
{\it Proof.}
We only have to prove that $\det \Phi\neq 0$.
This follows by checking the leading term for $\zeta\sim \infty$.
Using (2.48), (2.64a,b) and (2.67) we have
$$
\Phi(x,\zeta) \sim  \pmatrix{\phantom{-}1&\cdots&\phantom{-}1\cr
                  -\zeta_0&\cdots&-\zeta_{r-1}\cr
                   \vdots &\cdots&\vdots\cr
 (-\zeta_0)^{r-1}&\cdots&(-\zeta_{r-1})^{r-1}\cr}
\pmatrix{e^{\zeta_0Dx}\hat\phi_0(x)&{}&{}\cr
            {}&\ddots&{}\cr
            {}&{}&e^{\zeta_{r-1}Dx}\hat \phi_0(x)\cr}
\eqno(2.69)$$
for $\zeta\sim\infty$.
The determinant of the matrix on the right hand side of
equation (2.69) is non-zero. \quad $\square$
\medskip

 From Lemma 2.8 we obtain
$$
\Phi(x+2\pi,\zeta) = \Phi(x,\zeta)\Gamma(\zeta),
\eqno(2.70)$$
where
$$
\Gamma(\zeta) =
{\rm diag\,}\left(\gamma(\zeta_0), \ldots,\gamma(\zeta_{r-1})\right).
\eqno(2.71)$$
It follows that $T = \Phi(2\pi)\Phi(0)^{-1}$ is conjugate to $\Gamma$.
By combining this with (2.61), (2.55) and (2.46),
we arrive at the main result of this section.

\medskip
\noindent
{\bf Theorem 2.10.}
{\it The monodromy matrix of ${\cal L}$ is conjugate
to the diagonal matrix ${\cal T}$ given by
$$
{\cal T}=\exp
\pmatrix{
2\pi\zeta_0D-{1\over r}{\cal H}(\zeta_0)&{}&{}&{}\cr
{}&2\pi\zeta_1D-{1\over r}{\cal H}(\zeta_1)&{}&{}\cr
{}&{}&\ddots&{}\cr
{}&{}&{}&2\pi\zeta_{r-1}D-{1\over r}{\cal H}(\zeta_{r-1})\cr}\,,
\eqno(2.72)$$
where
$$
{\cal H}(\zeta)=
\int_0^{2\pi}(-D)^r [u_1]_{\rm diag} +
\sum_{k=1}^\infty (-\zeta)^{-k} {r\over k}
\int_0^{2\pi} \res ({\hat L}_1^{k/r})\,,
\eqno(2.73)$$
and $\hat L$ is the diagonalized form of the operator $L\in M$
corresponding to ${\cal L}\in {\cal M}_c$.}

\medskip
\noindent
{\bf Corollary 2.11.}
{\it All Hamiltonians of the reduced AKS system
carried by $M$ are generated by the ones in the following list:
$$
{\cal H}_{0,i}=(-1)^r\int_0^{2\pi} \left(D^r u_1\right)_{ii}\,,
\,\quad
{\cal H}_{k,i}={r\over k}\int_0^{2\pi}\res \left({\hat
L}_1^{k/r}\right)_{ii}\,,
\quad i=1,\ldots,p;\  k=1,2,\ldots\ .
\eqno(2.74)$$
The corresponding flows are subject to the relations
$$
\{ f\,,\,{\cal H}_{k,i}\}^{(2)}=\{ f\,,\,{\cal H}_{k+r,i}\}^{(1)}\,
\qquad \forall\,f\in C^\infty(M)\,,\quad \forall\, i,k.
\eqno(2.75)$$
The first Hamiltonians in each of the bihamiltonian ladders
belonging to fixed $k\  {\rm mod}\  r$ and fixed $i$ are
Casimirs of the first Gelfand-Dickey bracket,
$$
\{f\,,\,{\cal H}_{k,i}\}^{(1)}=0\,, \qquad
\forall\, f\in C^\infty(M) \quad\hbox{for}\quad k=0,1,\ldots,(r-1)\,.
\eqno(2.76)$$
The number of independent bihamiltonian ladders in (2.75)
is $pr-1=n-1$ since
$$
\sum_{i=1}^p {\cal H}_{mr,i}=0\,,
\qquad\hbox{\it for any}\qquad
m=1,2,\ldots\ ,
\eqno(2.77)$$
and $\sum_{i=1}^p {\cal H}_{0,i}$ is a Casimir with respect
to both Gelfand-Dickey Poisson brackets.}

\medskip
\noindent
{\it  Proof of Corollary 2.11.}
It is clear from the form of the matrix ${\cal T}$ in (2.72) that
the Hamiltonians of the reduced AKS system;
i.e., those obtained by reduction from  the
${\rm ad}^*$-invariant Hamiltonians on the dual of the
loop algebra,
are generated by those in the list (2.74).
The relations given by (2.75) and (2.76) can be traced back
to the ``general recursion relation''
given by (1.10),  but can also be verified directly by using
the expressions of the Gelfand-Dickey brackets,
(2.31), (2.32), and the Hamiltonians (2.74).
That $\sum_{i=1}^p {\cal H}_{0,i}$ is a Casimir of both Poisson
brackets also follows by direct verification.
As for (2.77), note that
$$
\sum_{i=1}^p {\cal H}_{mr,i}=\int_0^{2\pi}\tr \res {\hat L}^m=
\int_0^{2\pi}\tr \res L^m=0\,,
\eqno(2.78)$$
where the second equality follows from (2.44) and the
third equality holds because  $L$ is a differential
operator. \quad $\square$

\medskip
\noindent
{\it Remark 2.5.}
Let $\rho$ be a $p\times p$ diagonal matrix whose entries are
all $r$th roots of $1$,
$$
\rho={\rm diag\,}(\rho_1,\ldots,\rho_p)
\qquad\hbox{with}\qquad
(\rho_i)^r=1\,.
\eqno(2.79)$$
There is a unique $r$th root of $L$ whose leading term
is given by $\rho\Delta \pa$.
We denote  this pseudo-differential operator by $L^{1/r}_\rho$.
By using (2.43), (2.44) and (2.56) we can write
$$
L^{1/r}_\rho = g \rho \hat L_1^{1/r} g^{-1}\,,
\eqno(2.80)$$
which implies that
$$
{r\over k}\Tr (L^{k/r}_{\rho})={r\over k}\Tr (\rho^k {\hat L}_1^{k/r})=
\sum_{i=1}^p \rho_i^k {\cal H}_{k,i}\,.
\eqno(2.81)$$
For the Hamiltonian
${\cal H}_k^\rho := {r\over k}\Tr L^{k/r}_{\rho}$
($k\neq 0\  \hbox{mod}\  r$) we have
$$
\nabla_L {\cal H}_k^\rho= L_\rho^{k/r-1}\,.
\eqno(2.82)$$
  From this we obtain, as in the scalar case [1],
$$
\dot L = \{L\,,\,{\cal H}^\rho_k\}^{(2)}=
\{L\,,\,{\cal H}_{k+r}^\rho\}^{(1)}=[(L^{k/r}_{\rho})_+\,,\,L]\,.
\eqno(2.83)$$
There are $r^p$ possible choices of the matrix
$\rho$ in (2.79) and it is possible
to single out
$p$ of them in such a way that the corresponding ${\cal H}_k^\rho$'s
form a basis for the linear space spanned by the ${\cal H}_{k,i}$'s,
for any fixed $k\neq 0\ \hbox{mod}\ r$.
Thus for $k\neq 0\  \hbox{mod}\  r$
the relations (2.75) and (2.76) follow from (2.83).

\medskip
\noindent
{\it Remark 2.6.}
Note that the diagonal terms of the potential
$u_1$ of $L$ in (2.40b) are
constant along the flows of the hierarchy.
This is obvious for flows of the type (2.83), and
follows for all the flows
generated by the ${\rm ad}^*$-invariants by combining the
following two facts, which are easy to verify.
First, the function
$$
{\cal H}_K(L)=(-1)^r \int_0^{2\pi} \tr \left( K D u_1 D^{r-1}\right)\,,
\eqno(2.84)$$
where $K$ is any $gl_p$ valued function on $S^1$,
generates via the second Gelfand-Dickey bracket the
following one parameter group of transformations:
$$
f_t: L\mapsto e^{tK} L e^{-tK}\,.
\eqno(2.85)$$
Second, these transformations
leave the Hamiltonians (2.74) invariant for
arbitrary {\it diagonal} $K$.
For example, if $K={\rm diag\,}(K_1,\ldots,K_p)$
is a constant diagonal matrix then
$$
{\cal H}_K=\sum_{i=1}^p K_i {\cal H}_{0,i}
\eqno(2.86)$$
is one of the Hamiltonians of the hierarchy.
It generates the evolution equation
$$
\dot L =\{L\,,\,{\cal H}_K\}^{(2)}=[K,L]\,,
\eqno(2.87)$$
whose flow (2.85) indeed
leaves $[u_1]_{\rm diag}$ invariant (but not the full matrix $u_1$).
Finally, we could also consider symmetry reductions
of the hierarchy under the abelian group
action given by (2.85) with diagonal $K$'s.
The simplest reduction would be defined by setting $\tr (D^ru_1)=0$.
In this case, the reduced system
is the same as the one obtained by using
the Lie algebra $sl_n$ instead of
$gl_n$ throughout the construction.
Another symmetry reduction, which in a sense is maximal,
is obtained by setting $[u_1]_{\rm diag}=0$.
This leads to the system studied previously by
Gelfand-Dickey [15], Manin [16] and Wilson [17].

\bigskip
\noindent
{\bf 2.4. Some explicit formulae in the case $r=2$}
\medskip

Here we consider the simplest example $r=2$ in order
to make our general results more concrete.
We shall display the explicit form of the Poisson brackets
and the first two Hamiltonians for each bihamiltonian
ladder in our reduced AKS system.

Denote the general element of the phase space $M$ as
$$
L=\Delta^2\pa^2 + u\pa + w\,.
\eqno(2.88)$$
We can easily write out the evolution equations generated by
an arbitrary ${\cal H}\in C^\infty(M)$ via either of the
two Gelfand-Dickey Poisson brackets by using
$$
\nabla_L {\cal H}=\pa^{-2}\dhu +\pa^{-1}\dhw\,.
\eqno(2.89)$$

The Hamiltonian equation generated by means of the first Poisson
bracket (2.31) is given by
$$
\dot u=\{ u\,,\,{\cal H}\}^{(1)}=[\,\Delta^2\,,\,\dhw\, ]
\eqno(2.90{\rm a})$$
and
$$
\dot w=\{ w\,,\,{\cal H}\}^{(1)}=
\Delta^2 \left(\dhw\right)^\prime +\left(\dhw\right)^\prime \Delta^2
+[\Delta^2\,,\,\dhu\,]+[u\,,\,\dhw\,]\,.
\eqno(2.90{\rm b})$$
Observe that $[u]_{\rm diag}$ is in the centre
of the first Poisson bracket since it
does not change under any Hamiltonian flow.
In the scalar case $p=1$ all the commutator terms
drop out of (2.90), and for $u=0$ we would recover the first
Poisson structure of the standard KdV hierarchy.

The Hamiltonian equation  generated by ${\cal H}$ through
the second Poisson bracket (2.32) is given by
$$
\eqalign{
\dot u=\{u\,,\,{\cal H}\}^{(2)}=&
\left(\Delta^2\dhu u-u\dhu \Delta^2\right)+
\left(\Delta^2\dhw w-w\dhw \Delta^2\right)\cr
&-\Delta^2\left(\dhw u\right)^\prime
-2\Delta^2\left(\dhu\right)^\prime\Delta^2
+\Delta^2\left(\dhw\right)^{\prime\prime}\Delta^2\,,\cr}
\eqno(2.91{\rm a})$$
and
$$
\eqalign{
\dot w=\{w\,,\,{\cal H}\}^{(2)}=
&\Delta^2\left(\dhw w\right)^\prime
+w\left(\dhw\right)^\prime\Delta^2
+\Delta^2\left(\dhw\right)^{\prime\prime\prime}\Delta^2
-\Delta^2\left(\dhu\right)^{\prime\prime}\Delta^2
\cr
&+\left(\Delta^2\dhu w -w\dhu \Delta^2\right)
+\left(u\dhw w-w\dhw u\right)
-u\left(\dhw u\right)^\prime
\cr
&+u\left(\dhw\right)^{\prime\prime}\Delta^2
-\Delta^2\left(\dhw u\right)^{\prime\prime}
-u\left(\dhu\right)^\prime\Delta^2
\,.\cr}
\eqno(2.91{\rm b})$$

By straightforward computation from (2.43-45)
we obtain
$$
{\cal H}_{2,i}(u,w)=\int_0^{2\pi} \res (\hat L)_{ii}=
\int_0^{2\pi} h_{2,i}\,,
\eqno(2.92{\rm a})$$
with
$$
\eqalign{
h_{2,i}=&\sum_{k\neq i}
{{\Delta_i^2+\Delta_k^2}\over {(\Delta_i^2-\Delta_k^2)^2}}
u_{ik}u^\prime_{ki}
-\sum_{k\neq i}
{u_{ik}u_{ki}u_{ii}\over {(\Delta_i^2-\Delta_k^2)^2}}\cr
&+\sum_{k\neq i} \sum_{l\neq i} {{u_{ik}u_{kl}u_{li}}\over
{(\Delta_i^2-\Delta_k^2)(\Delta_i^2-\Delta_l^2)}}
+\sum_{k\neq i}{{u_{ik}w_{ki}+w_{ik}u_{ki}}\over
{(\Delta_i^2-\Delta_k^2)}}\,,\cr}
\eqno(2.92{\rm b})$$
where the $\Delta_i$'s are the components the
$p\times p$ diagonal matrix $\Delta$.
The above formulae may be checked by verifying that the
first evolution equation belonging to the bihamiltonian ladder
containing ${\cal H}_{2,i}$ can be written as
$$
\dot L=\{L\,,\,{\cal H}_{0,i}\}^{(2)}=\{L\,,\,{\cal H}_{2,i}\}^{(1)}
=[e_{ii}\,,\,L ]\,,
\qquad \left({\cal H}_{0,i}(u,w)=\int_0^{2\pi}
{u_{ii}\over \Delta_i^2}\right)\,,
\eqno(2.93)$$
consistently with (2.75), (2.86) and (2.87).
The equation generated by ${\cal H}_{2,i}$ via the second
Poisson bracket can now be obtained  by
substituting the expressions
for the functional derivatives ${\delta {\cal H}_{2,i}}\over {\delta u}$
and ${\delta {\cal H}_{2,i}}\over {\delta w}$
into (2.91a,b).
The result is somewhat lengthy, so we do not display it here.

We next give the first two Hamiltonians in the bihamiltonian
ladder of eq.~(2.83); i.e.,
$$
{\cal H}_k^\rho = \sum_{i=1}^p (\rho_i)^k {\cal H}_{k,i}=
 {2\over k}\int_0^{2\pi} \tr \res L^{k/2}_\rho
=\int_0^{2\pi} h_k^\rho
\qquad\hbox{for}\qquad k=1,3\,,
\eqno(2.94)$$
where the entries of $\rho={\rm diag\,}(\rho_1,\ldots,\rho_p)$
are
$\pm 1$'s, (see (2.79)).
To write down the result, it is useful to define
the diagonal matrix $\sigma$ by
$$
\sigma ={\rm diag\,}(\sigma_1,\ldots,\sigma_p)
:= \rho \Delta\,,
\eqno(2.95)$$
and to associate to any $p\times p$
matrix $v$ the matrix $\bar v$ given by the formula
$$
({\overline v})_{ij}:={v_{ij}\over (\sigma_i +\sigma_j)}\,.
\eqno(2.96)$$
Using this notation, we have
$$
h_1^\rho =\tr \sigma^{-1}w -\tr \sigma^{-1} \bu^2\,,
\eqno(2.97)$$
and
$$\eqalign{
h_3^\rho=&\tr \bw
\left(w-{4\over 3}\bu^\prime\sigma - 2\sigma \bu^\prime -2\bu^2\right)
+\tr \bu^\prime\left(\sigma{\overline \bu}^\prime\sigma
+{4\over 3}\sigma^2 {\overline \bu}^\prime\right)\cr
&+\tr \bu^2\left({\overline {(\bu^2)}}+2\sigma {\overline {\bu}}^\prime
+{4\over 3}{\overline {\bu}}^\prime \sigma^2\right)\,.\cr}
\eqno(2.98)$$
The first case of equation (2.83) is given explicitly as follows:
$$
\eqalignno{
\dot u &=\{ u\,,\,{\cal H}_1^\rho\}^{(2)}=\{u\,,\,{\cal H}_3^\rho\}^{(1)}
= \sigma u^\prime -2\sigma^2 \bu^\prime +[\bu\,,\,u]+[\sigma\,,\,w]\,,
&(2.99{\rm a}) \cr
\dot w &=\{ w\,,\,{\cal H}_1^\rho\}^{(2)}=\{w\,,\,{\cal H}_3^\rho\}^{(1)}
=\sigma w^\prime -\sigma^2 \bu^{\prime\prime} -u \bu^\prime
+[\bu\,,\,w]\,.
&(2.99{\rm b})}
$$
These equations simplify to the free chiral wave equation in
the scalar case $p=1$ after putting $u=0$, as it should.
The analogue of the KdV equation is generated by ${\cal H}_3^\rho$
through the second Poisson bracket (2.91a,b).
It is straightforward to derive it from the above,
but the final expression is quite long.

\medskip
\noindent
{\it Remark 2.7.}
There is no singularity
in the formulae (2.92b), (2.96-98) above since
the diagonal matrix $\Delta^2=(-D)^{-2}$ has distinct,
non-zero eigenvalues.
This goes back to choosing
$\Lambda=\Lambda_r\otimes D$ as a {\it regular} element
of the Heisenberg subalgebra at the beginning
of the construction.
In general, for any $r$,
if we continuously deform $D$ to the unit matrix,
which represents a singular case
(type II in the terminology of [7]),
then the Hamiltonians ${\cal H}_{mr,i}$ in (2.74) ($m=1,2,\ldots$)
may be expected to become $\infty$ and thus disappear from
the hierarchy.
In addition, it follows from a result in [16]
that in the $D=I$ case there exists only a single
$r$th root of $L$, up to a scalar multiple.

\bigskip
\bigskip


\vfill\eject

\centerline{\bf 3. Discussion}
\medskip

This paper was aimed at further exploring the integrable systems
that can be associated to graded regular elements of loop algebras
by the method of Drinfeld and Sokolov.
We have concentrated on the simplest case, given by the nontwisted
loop algebra $\ell(gl_n)$, and have
shown that graded regular elements exist only in those
Heisenberg subalgebras that correspond to partitions
into equal blocks $n=pr$ or equal blocks plus a singlet $n=pr+1$.
We further analyzed the first case by taking the grade $1$ regular
elements and proved that the generalized DS reduction results
in the matrix version of the $r$-KdV
hierarchy of Gelfand-Dickey.
We wish to close by mentioning some other interesting models, which
are related to this KdV type hierarchy in a way that is
familiar in the scalar case $p=1$.

First of all, one has the modified KdV type hierarchy
which is obtained from the general AKS system
by using the block diagonal gauge.
This means that the Hamiltonians of the modified
hierarchy are still given by the
functions of $L$ listed in (2.74), but $L$
(and consequently $\hat L$) is considered to depend on
$$
{\cal L}=\pa +\th +\La\,,
\qquad\hbox{where}\qquad
\th := {\rm diag\,}[\th_1,\ldots,\th_r]\,,
\eqno(3.1)$$
through the Miura map (2.22).
The evolution equations of the modified hierarchy
are generated by means of the free current algebra
Poisson bracket carried by the $\th_i$'s, which are
the basic fields of the modified hierarchy.

The nonabelian affine (periodic) Toda field equation may be
defined as follows.
Take the basic Toda field to be
a $Gl_n$-valued block diagonal matrix $g(x,t)$,
$$
g={\rm diag\,}[g_1,\ldots,g_r]\,,
\eqno(3.2)$$
where the $g_i$'s ($i=1,\ldots,r$) are $Gl_p$-valued functions,
periodic in the space variable $x$.
In addition to the grade $1$ regular element
$\La$, choose also a grade $-1$ regular element
$\bar \La$ and define the Toda equation to be
the zero curvature equation
$$
[{\cal L}_+ , {\cal L}_-]=0\,,
\eqno(3.3{\rm a})$$
where
$$
{\cal L}_+:= \pa_+ +g^{-1}\pa_+g  +\Lambda\,,
\qquad
{\cal L}_-:= \pa_-+g^{-1}\bar\La g\,,
\eqno(3.3{\rm b})$$
and $\pa_\pm = (\pa_x\pm \pa_t)$.
Here, ${\cal L}_+$ is obtained from
the modified KdV operator ${\cal L}$ in (3.1) by substituting
$\pa_+$ for $\pa$ and $g^{-1}\pa_+ g$ for $\th$, and
therefore we have the same
relationship between the
conservation laws of the present Toda model
and the modified KdV hierarchy as is familiar in the
abelian case [1], [10], [25].
In particular, by the same arguments as in [1],
we can construct an infinite number of conserved
local currents for the Toda equation
by transforming ${\cal L}_+$ into the Heisenberg subalgebra.
Note also that another infinite set of conserved currents
can be constructed by utilizing the fact that
the Toda equation can equivalently be written as
$$
[\tilde {\cal L}_+ , \tilde {\cal L}_-]=0\,,
\eqno(3.4{\rm a})$$
with
$$
\tilde {\cal L}_+:= \pa_+ +g\Lambda g^{-1}\,,
\qquad
\tilde {\cal L}_-:= \pa_--\pa_-g\cdot g^{-1} +\bar\La \,,
\eqno(3.4{\rm b})$$
and transforming $\tilde {\cal L}_-$ into the
Heisenberg subalgebra.

The nonabelian affine Toda model defined above
was proposed originally by Mikhailov [26]
and further studied, e.g.,   in [27].
More precisely, these authors took
$\La=\La_r\otimes 1_p$ and
$\bar \La =\La_r^{-1}\otimes 1_p$,
where $1_p$ is the $p\times p$ unit matrix.
These are not regular elements of the Heisenberg
subalgebra and therefore the DS construction
of local conservation laws corresponding to
graded generators of the Heisenberg subalgebra
would not be applicable with this choice without
modifications.

A related nonabelian conformal (open) Toda model
can be obtained by omitting the $\lambda$ dependent terms
from ${\cal L}_\pm$ ($\tilde {\cal L}_\pm$) in the above.
This model is a member of a family of models described
in [28] and can also be obtained by
a Hamiltonian symmetry reduction of the
Wess-Zumino-Novikov-Witten (WZNW) model.
By using the WZNW picture it is clear that the
${\cal W}$-algebra given by the second
matrix Gelfand-Dickey Poisson bracket can be realized as
the algebra of Noether currents in the nonabelian conformal
Toda model.
Details can be found in refs.~[4], [5], [18].
Note that in these papers the group $Sl_n$
was used instead of $Gl_n$, but
one can impose the $Sl_n$ constraints
$\tr (D^r u_1)=0$, $\tr \th=0$ and ${\rm det\,}g=1$ without
changing the essential features of the KdV,
modified KdV or Toda systems,
or their relationship.

It seems plausible that the nonabelian
affine Toda model (as well as an appropriate
``conformal affine'' variant of it) can be viewed as
a Hamiltonian symmetry reduction of the affine
WZNW model, generalizing the abelian case
[29].
This Toda model is also known to be
a reduction of the multicomponent Toda lattice hierarchy [30].
Similarly, the matrix $r$-KdV hierarchy should be related to
the multicomponent KP hierarchy [30], [31].

The other case
in which a graded regular element exists in the
Heisenberg subalgebra, corresponding to the partition
$n=pr+1$,
has not been pursued in the present work.
By taking an arbitrary regular element of minimal positive
grade it may be verified that the generalized DS reduction
proposed in [7] leads to a ${\cal W}$-algebra
which is again equal to one of those studied in [4], [5]
in the context of WZNW reductions.
In these papers a family
of ${\cal W}$-algebras was associated to
the $sl_2$ subalgebras of $gl_n$ ($sl_n$).
The  ${\cal W}$-algebra arising
from a regular element of minimal grade corresponds to the
$sl_2$ subalgebra under which the defining
representation of $gl_n$ decomposes into $p$ copies of
the $r$-dimensional $sl(2)$ irreducible representation
plus a singlet.

Recall that both the $sl_2$ subalgebras of $gl_n$
and the Heisenberg subalgebras of $\ell(gl_n)$
are classified by the partitions of $n$.
It is unclear whether there is a general relationship
between all ${\cal W}$-algebras associated to
$sl_2$ embeddings and KdV type hierarchies
or not, since there is a ${\cal W}$-algebra
for any partition, but graded regular elements
exist only in exceptional cases.

The present work was based entirely on the Hamiltonian
AKS approach to integrable systems.
The Grassmannian approach [32], [24]
has also been generalized to other Heisenberg
subalgebras than the principal one in [33], [31], [9].
It is clear that the
integrable systems associated to
graded regular elements in arbitrary loop algebras would
deserve further study from both viewpoints.
The starting point could be the explicit description
of Heisenberg subalgebras recently worked out in [34].

 \bigskip\bigskip
\noindent
{{\bf Acknowledgements.\ }
This work was supported in part by the Natural Sciences and Engineering
Research Council of Canada and the Fonds FCAR du Qu\'ebec. One of
us (L.F.) wishes to thank L. Vinet and the Laboratoire de physique nucl\'eaire,
Universit\'e de Montr\'eal for hospitality during the course of this work.}
\bigskip\bigskip

\vfill\eject


\centerline{\bf References}
\medskip

\item{[1]}
Drinfeld, V.~G., Sokolov, V.~V.:
Lie algebras and Equations of Korteweg - De Vries Type.
Jour. Sov. Math. {\bf 30} 1975-2036 (1985);
Equations of Korteweg-De Vries type and Simple Lie Algebras.
Soviet. Math. Dokl. {\bf 23} 457-462 (1981)

\item{[2]}
Douglas, M.:
Strings in less than one-dimension and the generalized KdV hierarchies.
Phys. Lett. {\bf 238B} 176-180 (1990)

\item{[3]}
Zamolodchikov, A.~B.:
Infinite additional symmetries in two dimensional CFT.
Theor. Math. Phys. {\bf 65} 1205-1213 (1988);
\item{}
Lukyanov, S.~L., Fateev, V.~A.:
Additional symmetries and exactly soluble models in two dimensional
conformal field theory.
Kiev preprints ITF-88-74R, ITF-88-75R, ITF-88-76R (1988)

\item{[4]}
Bais, F.~A., Tjin, T., van Driel, P.:
Covariantly coupled chiral algebras.
Nucl. Phys. {\bf B357} 632-654 (1991)

\item{[5]}
Feh\'er, L., O'Raifeartaigh, L., Ruelle, P., Tsutsui, I., Wipf, A.:
On Hamiltonian reductions of the Wess-Zumino-Novikov-Witten theories.
Phys. Rep. (in press);
Generalized Toda Theories and ${\cal W}$-Algebras Associated
with Integral Gradings. Ann. Phys. (N.~Y.) {\bf 213} 1-20 (1992)

\item{[6]}
Frenkel, E., Kac, V.~G. , Wakimoto, M.,
Characters and Fusion Rules for W-Algebras via Quantized Drinfeld-Sokolov
Reduction.
Commun. Math. Phys. {\bf 147} 295-328 (1992)
\item{}
Bowcock, P., Watts, G.~M.~T.: On the classification of quantum W-algebras.
Nucl. Phys. {\bf B379} 63-96 (1992);
\item{}
Gervais, J.L., Matsuo, Y.:
Classical $A_n$--W-Geometry.
Paris preprint LPTENS-91/35 (1992)
\item{}
Frappat, L., Ragoucy, E., Sorba, P.:
W-algebras and superalgebras from constrained WZW models:
a group theoretical classification.
Lyon
preprint ENSLAPP-AL-391/92 (1992)

\item{[7]}
De Groot, M.~F., Hollowood, T.~J., Miramontes, J.~L.:
Generalized Drinfeld-Sokolov Hierarchies.
Commun. Math. Phys. {\bf 145} 57-84 (1992)
\item{}
Burroughs, N.~J., De Groot, M.~F., Hollowood, T.~J., Miramontes, J.~L.:
Generalized Drinfeld-Sokolov Hierachies II: The Hamiltonian Structures.
Princeton preprint PUTP-1263, IASSN-HEP-91/42 (1991);
Generalized W-algebras and integrable hierarchies.
Phys. Lett. {\bf 277B} 89-94 (1992)

\item{[8]}
Burroughs, N.~J.:
Co-adjoint orbits of the generalised $Sl(2)$, $Sl(3)$ KdV
hierarchies.
Nucl. Phys. {\bf B379} 340-376 (1992)

\item{[9]}
Hollowood, T.~J., Miramontes, J.~L.:
Tau-Functions and Generalized Integrable Hierarchies.
preprint OUTP-92-15P, CERN-TH-6594/92 (1992)

\item{[10]}
Wilson, G.~W.:
The modified Lax and two-dimensional Toda lattice equations
associated with simple Lie algebras.
Ergod. Th. and Dynam. Sys. {\bf 1} 361-380 (1981)

\item{[11]}
McIntosh, I.:
Groups of equivariant loops in SL(n+1,C) and soliton equations.
Newcastle preprint (1992)

\item{[12]}
Kac, V.~G., Peterson, D.~H.:
112 constructions of the basic representation of the loop group of $E_8$.
In: Proceedings of Symposium on Anomalies, Geometry and Topology.
Bardeen, W.~A., White, A.~R. (eds.)
Singapore: World Scientific 1985

\item{[13]}
ten Kroode, F., van de Leur, J.:
Bosonic and fermionic realizations of the affine algebra $\widehat{gl}_n$.
Commun. Math. Phys. {\bf 137} 67-107 (1991)

\item{[14]}
Dickey, L.~A.:
Soliton Equations and Hamiltonian Systems.
Adv. Ser. in Math. Phys. Vol. 12.
Singapore: World Scientific 1991

\item{[15]}
Gelfand, I.~M., Dickey, L.~A.:
The resolvent and Hamiltonian systems.
Funct. Anal. Appl. {\bf 11} 93-104 (1977)

\item{[16]}
Manin, Yu.~I.:
Algebraic aspects of nonlinear differential equations.
Jour. Sov. Math. {\bf 11} 1-122 (1979)

\item{[17]}
Wilson, G.:
Commuting flows and conservation laws for Lax equations.
Math. Proc. Cambr. Philos. Soc. {\bf 86} 131-143 (1979);
On two constructions of conservation laws for Lax equations.
Q. J. Math. Oxford {\bf 32} 491-512 (1981)

\item{[18]}
Tjin, T., van Driel, P.:
Coupled WZNW-Toda models and Covariant KdV hierarchies.
Amsterdam preprint IFTA-91-04 (1991)

\item{[19]}
Reyman, A.~G., Semenov-Tian-Shansky, A.:
Current algebras and nonlinear partial differential equations.
Soviet. Math. Dokl. {\bf 21} 630-634 (1980);
A family of Hamiltonian structures, hierarchy of Hamiltonians,
and reduction for first-order matrix differential operators.
Funct. Anal. Appl. {\bf 14} 146-148 (1980)

\item{[20]}
Semenov-Tian-Shansky, M.~A.:
What is a classical r-matrix.
Funct. Anal. Appl. {\bf 17} 259-272 (1983);
Dressing Transformations and Poisson Group Actions.
Publ. RIMS Kyoto Univ. {\bf 21} 1237-1260 (1985)

\item{[21]}
Reyman, A.~G., Semenov-Tian-Shansky, A.:
Compatible Poisson structures for Lax equations: an r-matrix
approach.
Phys. Lett. {\bf A130} 456-460 (1988)

\item{[22]}
Faddeev, L.~D., Takhtajan, L.~A.:
Hamiltonian Methods in the Theory of Solitons.
Heidelberg: Springer-Verlag 1986

\item{[23]}
Kupershmidt, B.~A., Wilson, G.:
Modifying Lax equations and the second Hamiltonian structure.
Invent. Math. {\bf 62} 403-436 (1981)

\item{[24]}
Segal, G., Wilson, G.:
Loop groups and equations of KdV type.
Publ. Math. IHES 5-65 (1985)

\item{[25]}
Mikhailov, A.~V., Olshanetsky, M.~A., Perelomov, A.~M.:
Two-dimensional generalized Toda lattice.
Commun. Math. Phys. {\bf 79} 473-488 (1981)
\item{}
Kupershmidt, B.~A., Wilson, G.:
Conservation Laws and Symmetries of Generalized Sine-Gordon Equations.
Commun. Math. Phys. {\bf 81} 189-202 (1981)

\item{[26]}
Mikhailov, A.~V.:
The reduction problem and the inverse scattering method.
Physica {\bf 3D} 73-117 (1981)

\item{[27]}
Harnad, J., Saint-Aubin, Y., Shnider, S.:
The Soliton Correlation Matrix and the Reduction Problem for
Integrable Systems.
Commun. Math. Phys. {\bf 93} 33-56 (1984)

\item{[28]}
Leznov, A.~N., Saveliev, M.~V.:
Group-Theoretical Methods for Integration of Nonlinear
Dynamical Systems.
Progress in Physics Vol.~15.
Basel: Birk\"auser Verlag 1992

\item{[29]}
Babelon, O., Bonora, L.:
Conformal affine $sl_2$ Toda field theory.
Phys. Lett. {\bf B244} 220-226 (1990)
\item{}
Aratyn, H., Ferreira, L. A., Gomes, J. F., Zimerman, A. H.:
Kac-Moody construction of Toda type field theories.
Phys. Lett. {\bf B254} 372-380 (1991)

\item{[30]}
Ueno, K., Takasaki, K.: Toda lattice hierarchy.
in: Advanced Studies in Pure Mathematics {\bf 4} 1-95 (1984)

\item{[31]}
Dickey, L.~A.:
Another example of a $\tau$-function.
In: Proceedings of the CRM Workshop on Hamiltonian Systems,
Transformation Groups and Spectral Transform Methods.
Harnad, J., Marsden, J.~E. (eds.)
Montr\'eal: CRM publications 1990.

\item{[32]}
Date, E., Kashiwara, M., Jimbo, M., Miwa, T.:
Transformation Groups for Soliton Equations.
In: Non-linear Integrable Systems -- Classical Theory and
Quantum Theory. Proceedings of the RIMS Symposium.
Jimbo, M., Miwa, T. (eds.) Singapore: World Scientific 1983

\item{[33]}
Bergvelt, M.~J., ten Kroode, A.~P.~E.:
$\tau$-functions and
zero curvature equations of Toda-AKNS type.
J. Math. Phys. {\bf 29} 1308-1320 (1988)
\item{}
Kac, V.~G., Wakimoto, M.:
Exceptional Hierarchies of Soliton Equations.
Proceedings of Symposia in Pure Mathematics {\bf 49} 191-237 (1989)
\item{}
Dickey, L.~A.:
On the $\tau$-function of matrix hierarchies of integrable equations.
J. Math. Phys. {\bf 32} 2996-3002 (1991)
\item{}
van Driel, P.:
On a new hierarchy associated with the $W_3^{(2)}$ conformal algebra.
Phys. Lett. {\bf B274} 179-185 (1992)
\item{}
Adams, M.~R., Bergvelt, M.~J.:
The Krichever map, vector bundles over algebraic curves, and
Heisenberg algebras. Univ. Illinois preprint (1991)

\item{[34]}
ten Kroode, F., van de Leur, J.:
Bosonic and fermionic realizations of the affine algebra $\widehat{so}_{2n}$.
Utrecht preprint no. 636 (1991);
Level one representations of the affine Lie algebra $B_n^{(1)}$.
Utrecht preprint no. 698 (1991);
Level one representations of the twisted affine algebras
$A_n^{(2)}$ and $B_n^{(2)}$.
Utrect preprint no. 707 (1992)

\vfill\eject

\bye